\documentclass[modern]{aastex63}

\usepackage{xspace}
\usepackage{cancel}

\definecolor{linkcolor}{rgb}{0.1216,0.4667,0.7059}

\usepackage{xcolor, fontawesome}
\definecolor{twitterblue}{RGB}{64,153,255}
\newcommand\twitter[1]{\href{https://twitter.com/#1 }{\textcolor{twitterblue}{\faTwitter}\,\tt \textcolor{twitterblue}{@#1}}}
\usepackage{amsmath}

\graphicspath{{../notebooks/}}

\shorttitle{Kernel Phase and Coronagraphy with Automatic Differentiation}
\shortauthors{B. J. S. Pope et al.}

\begin{document}

\title{Kernel Phase and Coronagraphy with Automatic Differentiation}

\correspondingauthor{Benjamin J. S. Pope \twitter{fringetracker}}
\email{benjamin.pope@nyu}

\author[0000-0003-2595-9114]{Benjamin J. S. Pope}
\affiliation{Center for Cosmology and Particle Physics, Department of Physics, New York University, 726 Broadway, New York, NY 10003, USA}
\affiliation{Center for Data Science, New York University, 60 Fifth Ave, New York, NY 10011, USA}
\affiliation{NASA Sagan Fellow}

\author[0000-0003-3818-408X]{Laurent Pueyo}
\affiliation{Space Telescope Science Institute, 3700 San Martin Drive, Baltimore, MD, 21218, USA}

\author[0000-0002-6171-9081]{Yinzi Xin}
\affiliation{Department of Aeronautics and Astronautics, Massachusetts Institute of Technology, Cambridge, MA 02139, USA}

\author[0000-0001-7026-6291]{Peter~G.~Tuthill}
\affiliation{Sydney Institute for Astronomy (SIfA), School of Physics, The University of Sydney, NSW 2006, Australia}

\begin{abstract}
The accumulation of aberrations along the optical path in a telescope produces distortions and speckles in the resulting images, limiting the performance of cameras at high angular resolution. It is important to achieve the highest possible sensitivity to faint sources, using both hardware and data analysis software. While analytic methods are efficient, real systems are better modelled numerically, but numerical models of complicated optical systems with many parameters can be hard to understand, optimize and apply.
Automatic differentiation or `backpropagation' software developed for machine learning applications now makes calculating derivatives with respect to aberrations in arbitrary planes straightforward for any optical system.
We apply this powerful new tool to the problem of high-angular-resolution astronomical imaging. Self-calibrating observables such as the `closure phase' or `bispectrum' have been widely used in optical and radio astronomy to mitigate optical aberrations and achieve high fidelity imagery. Kernel phases are a generalization of closure phases valid in the limit of small phase errors.
Using automatic differentiation, we reproduce existing kernel phase theory within this framework and demonstrate an extension to the case of a Lyot coronagraph, which is found to have self-calibrating combinations of speckles which are resistant to phase noise, but only in the very high wavefront quality regime. 
As an illustrative example, we reanalyze Palomar adaptive optics observations of the binary $\alpha$~Ophiuchi, finding consistency between the new pipeline and the existing standard.
We present a new Python package \textsc{morphine} that incorporates these ideas, with an interface similar to the popular package \textsc{poppy}, for optical simulation with automatic differentiation. These methods may be useful for designing improved astronomical optical systems by gradient descent.
\href{https://github.com/benjaminpope/morphine}{\color{linkcolor}\faGithub} 

\end{abstract}


\section{Introduction} 
\label{sec:intro}
Many questions in astronomy and other sciences can only be answered with diffraction-limited high resolution imaging. The highest resolutions are typically achieved with the method of interferometry, in which waves detected at multiple receivers are combined physically or in post-processing to obtain the Fourier transform of the source intensity distribution \citep{vc34,zernike38}. Even in the case of a single telescope or camera, it is often nevertheless helpful to think of them as an interferometer composed of many sub-apertures which combine their signals directly onto a focal plane \citep[a `Fizeau interferometer':][]{fizeau1868}. In all of these cases, unknown path delays through each receiver or subaperture are the dominant source of aberration with low spatial frequency structure causing issues such as varying `tip-tilt' of the final image position, while higher order distortions yield clouds of `speckles'.

Strategies to correct for these aberrations include the use of active systems such as adaptive optics or delay lines; calibrating these errors with comparison to a reference star; or self-calibration, using the physics of the noise process to correct for it in post-processing.
In the recovery of object phase information, among the most longstanding techniques is that of self-calibration with `closure phases' \citep[introduced in the context of radio astronomy by][]{jennison58}, in which phases are summed around three interferometric baselines which form a closing triangle. The phase error terms local to each subaperture cancel, so that three low-signal-to-noise (SNR) baseline phases deliver one high-SNR observable. In a non-redundant array (one in which no baseline vector is repeated between different pairs of subapertures) of sufficient size, a large number of closure phases can be obtained which anchor image reconstruction with great precision \citep{chael18}. `Closure amplitudes' which are resistant to fluctuations in input amplitude or gain can also be obtained by a similar construction using four telescopes \citep{twiss60,blackburn20}. The closure phase is the argument of a quantity called the triple product or `bispectrum', and it has recently been shown from the perspective of invariant theory in algebraic geometry that for a wide class of problems limited by phase noise, knowledge of the mean of a signal, its power spectrum, and its bispectrum are necessary and sufficient for an optimal signal reconstruction \citep{bandeira17}.

On the other hand, for the case of direct imaging of high contrast companions with coronagraphs, analytic self calibrations such as those above are not yet known, and external calibration is necessary. The standard approaches to data analysis in coronagraphy rely on exploring a diversity of PSFs experimentally, constructing a basis covering some of the diversity in speckle patterns, and then subtracting out a linear combination of vectors in this basis from measured data \citep[e.g. KLIP;][]{lafreniere07,soummer12,pueyo16}. Advantages can also be gained from angular differential imaging \citep[ADI;][]{marois06} and spectral deconvolution with wavelength \citep{sparks02}. In this paper we will extend our understanding of analytic self calibration to better include coronagraphs, so that we can add an additional layer of precision to exoplanet imaging calibration.

\subsection{Kernel Phases}

The kernel phase method is a way of extending closure phase from simple nonredundant arrays of telescopes to the densely-filled pupils of real telescopes. If we describe the propagation of phase and amplitude noise in terms of a matrix \citep{lannes1991}, we can obtain powerful generalizations to closure phases and amplitudes. We will describe the effect of aberrations on baseline phases as

\begin{equation}
    \mathbf{\Phi} = \mathbf{\Phi}_0 + \mathbf{A}_\phi \cdot \phi,
\end{equation}

\noindent where $\mathbf{\Phi}$ is a vector of observed phases on each baseline in the $u,v$ (focal plane Fourier domain), $\mathbf{\Phi}_0$ the true astrophysical phases, $\phi$ the phase noise at each point in the pupil plane, and $\mathbf{A}_\phi$ a transfer matrix from the pupil plane to the $u,v$ plane phases. A similar expression can be written for amplitudes \citep{pope16}. In the redundant pupil case such as a standard full telescope pupil, where each baseline appears more than once, the propagation of phase noise from the pupil to baselines is no longer linear, but in the limit where aberrations are small it can be linearized. For sufficiently small phase perturbations ($<< 1~\text{rad}$), error propagation to the focal plane Fourier phases is approximately linear: this is simply a Taylor expansion to first order of the nonlinear phase transfer operation, in which $\mathbf{A}_\phi$ is the Jacobian matrix of partial derivatives $\partial\Phi_j/\partial\phi_k$. For a single focal plane imaging system this operator $\mathbf{A}_\phi$ can be determined analytically for a given discrete pupil model. 

\citet{martinache10} introduced the idea of `kernel phase' to generalize closure phase to a redundant aperture. While in this case closure phases no longer exactly cancel out the contributions of aberrations $\phi$, nevertheless there is an equivalent that can be found numerically. A left kernel operator $\mathbf{K}$ can be calculated via singular value decomposition (SVD) that annihilates $\mathbf{A}_\phi$, such that

\begin{equation}
        \mathbf{K}\cdot\mathbf{A}_\phi = 0
\end{equation}

\noindent and therefore we can find self-calibrating kernel phases $\mathbf{K}\cdot\mathbf{\Phi}$ which are immune to phase noise to linear order:

\begin{align}
    \mathbf{K}\cdot\mathbf{\Phi} &= \mathbf{K}\cdot\mathbf{\Phi}_0 + \cancel{\mathbf{K}\cdot\mathbf{A}_\phi\cdot\phi} \\
    &= \mathbf{K}\cdot\mathbf{\Phi}_0.
\end{align}

Meanwhile, the phases in the complementary space to the kernel space have the opposite property, that they are especially sensitive to input phase aberrations, and for appropriate apertures can be used for wavefront sensing from the image domain \citep{martinache13,pope14,martinache16b}.

The kernel phase method has been applied to space-based data from the \textit{Hubble Space Telescope} NICMOS camera \citep{pope13,laugier19,martinache20}; ground-based data from the Palomar 200-Inch adaptive optics (AO) equipped Pharo camera \citep{palomar,martinache20}, the Large Binocular Telescope \citep{sallum15}, the VLT/NACO camera \citep{kammerer19}, MagAO \citep{sallum19b}, and explored theoretically for ground- and space-based telescopes \citep{ireland13,martinache11,sallum19a,ceau19}.

The kernel phase method relies on a Taylor expansion of the optical propagation, which for a simple telescope is easy to do analytically. But many systems in reality cannot be treated this way - for example, problems involving diffraction between multiple planes like in coronagraphs. We shall see in the following that automatic differentiation can supply derivatives for arbitrary numerically-simulated imaging systems, extending self-calibration to a wider class of instruments and also offering new opportunities for optical design and optimization.

\subsection{Automatic Differentiation}
The practical use of neural networks in machine learning applications is dependent on the efficient calculation of analytic gradients of often very complicated composite functions, for example the matrix operations composed with nonlinear activation functions that are seen in neural networks \citep{lecun15}. This problem is usually referred to as algorithmic differentiation, automatic differentiation or `autodiff', and is solved simply by the chain rule.
Autodiff is available in many implementations, such as the Python packages \textsc{TensorFlow} \citep{tensorflow2015}, \textsc{theano} \citep{theano}, \textsc{PyTorch} \citep{pytorch}, \textsc{autograd} \citep{autograd}, \textsc{jax} \citep{jax}, and many packages in the Julia language \citep{julia}. The reverse-mode autodiff or `backpropagation' algorithm \citep{linnainmaa1970,lecun1988theoretical} has made this practical for neural networks, but it is typically most-useful in cases where the output dimensionality is much smaller than the input, as for most neural networks. Forwards-mode autodiff, on the other hand, is usually better in the case where the output is of higher dimension than the input. For optics problems, both cases can be found -- for example, for optimization, backpropagation is usually the best approach, while for kernel phase analysis, forwards mode is better suited.

In simulating physical optics, we normally consider the complex electric field on a 2D plane, which is pixelized and then flattened to a 1D vector.
Optical propagation then consists of a series of Fourier (or  Fresnel) transforms mapping between planes, and matrix or element-wise multiplications by phase and amplitude screens in those planes. Indeed, for practical purposes in astronomy and imaging science generally, optical propagation through a whole system is generally a linear operation that could be written as a single (large) matrix multiplication. However quantities of interest are usually not specified as real and imaginary electric field components, but rather as amplitudes and phases, so that the relations between (for example) input and output phases is in general nonlinear. Finding the derivatives of $u,v$ phase with respect to pupil aberrations is therefore an ideal problem for autodiff.

The analogy between the operations of optics and deep neural networks is so strong that not only have autodiff packages been used to simulate photonic systems \citep[e.g.][]{hughes18}, photonic systems have in fact been used as analog computers implementing neural networks for machine learning \citep{hughes19,guo19}. 

Several groups have applied autodiff to areas of optical science relevant to astronomy. Autodiff has been applied fruitfully to geometric optics or ray-tracing \citep[e.g.][]{werner2012,sutin16}, which is important in astronomy for understanding gravitationally-lensed systems. \citet{chianese19} and \citet{morningstar19} have applied this method to integrating differentiable forwards models of gravitational lensing with neural networks for image analysis. Autodiff methods have also been applied for image reconstruction from interferometric data, including gravitationally-lensed systems \citep{morningstar18} and protoplanetary disks \citep{czekala19}.

The \textsc{DeepOptics} project \citep{sitzmann2018} has used autodiff to optimize `computational cameras'. Building their model in \textsc{TensorFlow}, they couple a physical optics simulation, detector simulation, and a deconvolution post-processing stage for a total end-to-end imaging simulation. Where normally you might optimize some heuristic of the PSF (such as the full width at half maximum), this makes it possible to jointly optimize hardware and software with respect to figures of merit of the overall system such as final resolution or depth of field. Some designs arrived at in this way are exotic: for example, super-resolution is achieved by finding a lens with three off-centre Fresnel-lens components focusing to three separate spots, which the deconvolution stage shifts and stacks. Because it is built in \textsc{TensorFlow}, the diffraction simulation can be incorporated as a `physical layer' in neural network applications in microscopy, for example for optimizing hardware and software for image classification \citep{muthumbi19}, or with reinforcement learning for adaptively learning sample illumination \citet{chaware19}.

An approach similar to \textsc{DeepOptics} is likely to be extremely valuable in designing, for instance, pupil masks for coronagraphy \citep[e.g.][]{guyon03,carlotti11} or for diffractive-pupil astrometry \citep[e.g.][]{guyon12,tuthill18}. 

Close to the topic of this paper, autodiff has been applied to the problem of phase retrieval \citep{jurling14,paine19}, inferring a wavefront from a PSF. In this context, a key advance of autodiff over previous methods is that we can trivially account for pixel sampling/binning and detector nonlinearity. These will be important issues when considering that the JWST mid-infrared imager MIRI will have significant detector nonlinearity \citep{rieke15}, or where we may wish to look at saturated sources. While they do not address more complex optical systems, this method may be straightforwardly applicable to sensing non-common-path errors in coronagraphic images. It may also be helpful to integrate phase retrieval simultaneously with flat-field calibration and light curve extraction in photometric missions such as \textit{Kepler}, K2, TESS and their successors, combining the gradient descent flat field optimization in halo photometry for bright stars \citep{white17,pope19} and PSF optimization for fainter stars in crowded fields \citep[e.g.][]{eleanor,nardiello19}. 

While \citet{jurling14} derive analytic expressions for many optically-relevant operations, machine-learning software that has become available since then have significantly widened the range of options and introduced more user-friendly APIs. These approaches have been taken up outside of optical imaging, for example in X-ray coherent diffractive imaging \citep{kandel19,nashed19} and nanotomography \citep{Dueaay3700}.

Rather than using autodiff gradients for optimization, in this paper we use them to understand optical systems and the information they propagate.
In the following we will show how autodiff can reproduce the existing state of the art in kernel phase, and demonstrate a way forward using autodiff to extend the kernel phase idea to coronagraphic instruments.

\section{Simulations}
\label{sec:method}

To forwards-model systems, we adapt the popular physical optics library \textsc{poppy} \citep{poppy}, with the goal of compatibility with existing simulations built on \textsc{poppy} such as \textsc{WebbPSF} \citep{webbpsf}. We use the matrix Fourier transform mode and avoid FFTs \citep{soummer07} to more easily generate arbitrary image sampling and for consistency with the kernel phase code \textsc{xara} \citep{martinache20}. We have adapted the low-level features of \textsc{poppy} to use the Google autodiff library \textsc{jax} \citep{jax} in place of \textsc{NumPy}, and to distinguish it from the original version we call this new \textsc{poppy} derivatives library \textsc{morphine}. 

This has several advantages already over analytic methods for kernel phase. \textsc{morphine} can calculate monochromatic or polychromatic PSFs: the polychromatic capability allows us to explicitly construct broadband kernel phase operators. It is also possible to take derivatives with respect to wavefronts specified in bases other than the pixel basis, for example Zernike or hexike modes, as we shall discuss in Section~\ref{zernike}.

\subsection{Simple Diffraction Example}
\label{sec:simple}

First we want to ensure basic kernel phase calculations are reproduced in this new model. We propagate monochromatic 2.0\,$\mu$m light through a 2.0\,m diameter circular pupil that is imaged onto a detector with a 20\,mas/pixel scale and 4\,arcsec field of view. The pupil plane and the baseline $u,v$ plane are each calculated on a 64x64 grid. We evaluate the Jacobian of the $u,v$ phases with respect to the input pupil phases using forwards-mode autodiff and then evaluate this Jacobian at zero phase to get our $\mathbf{A_\phi}$ matrix. On a laptop computer this takes a few minutes processing time. Memory usage is the main overhead: with 16GB RAM, grid size is limited to about 128x128. Demand on computational resources presents a major limitation to widespread application; in future this may be overcome by more efficient refactoring of the code, or simply by doing calculations on clusters with very large RAM allocations.

Although it is inefficient to calculate this full Jacobian with respect to a densely-pixelized basis for the wavefront in the pupil plane, we do so here for completeness. In practical cases, aberrations are typically dominated by low-order modes and we could consider the Jacobian only with respect to (for example) some much smaller number of Zernike modes, or pixels in the $u,v$ plane. Reducing the input dimensionality like this would have significant advantages in memory usage for forwards-mode autodiff.

\begin{figure}
\plotone{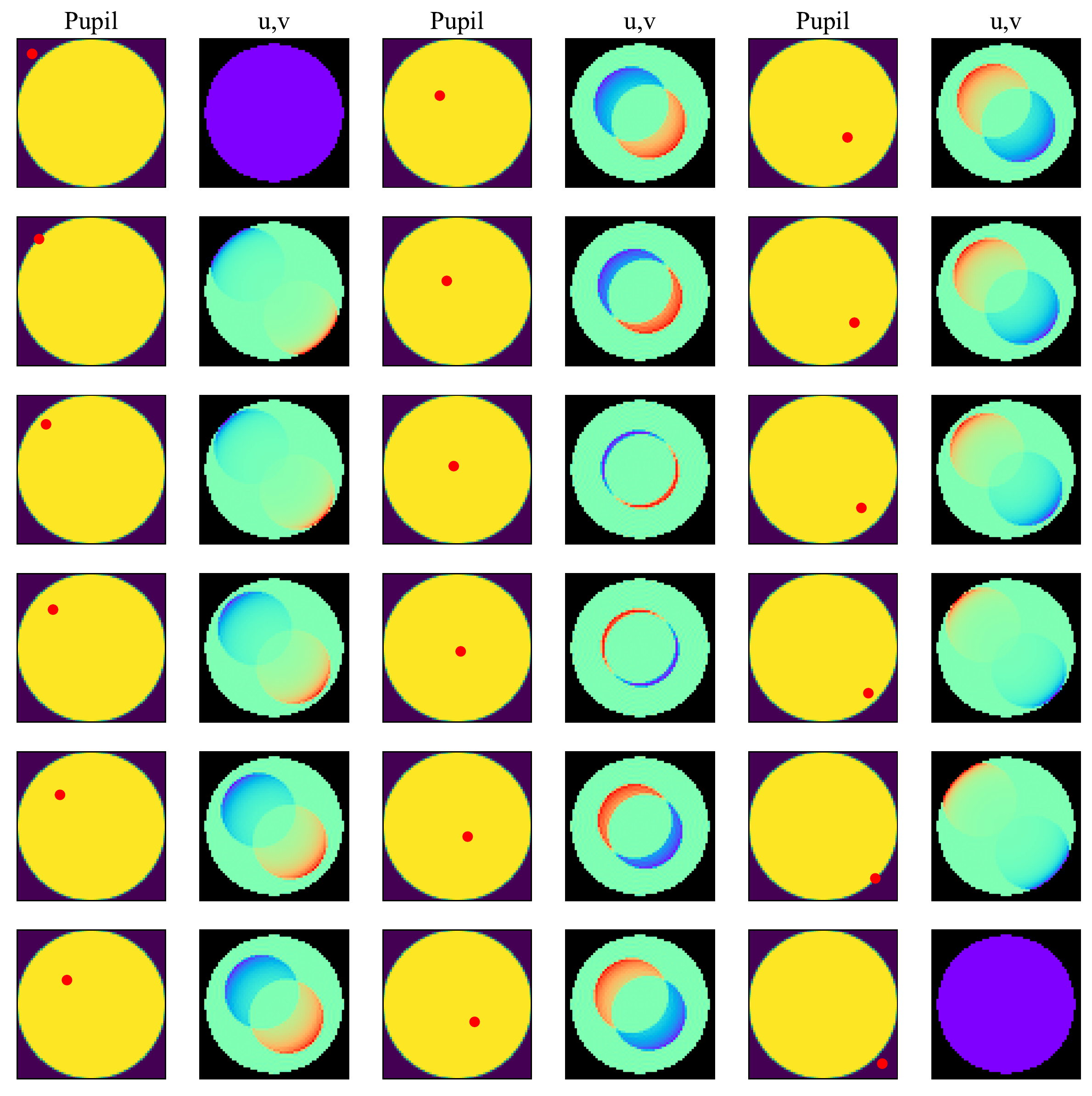}
\caption{A representation of the Jacobian determined by \textsc{jax} for monochromatic 2$\mu$m diffraction from a 2.0\,m diameter circular pupil. In each pair of images, on the left is the pupil, with a red dot highlighting a selected pixel; on the right is a map of the derivative of the phase in the $u,v$ plane with respect to the phase at that point in the pupil. These Jacobian vectors are pairs of rows and columns of the $\mathbf{A}$ phase transfer matrix. We see the same maps as in Figure~1 of \citet{martinache10}. \label{kernel_jacobian}}
\end{figure}

In Figure~\ref{kernel_jacobian} we display the results of our Jacobian calculation for a set of highlighted pupil samples and illustrating the influence of phase in the $u,v$ plane. 
The results are similar to Figure~1 of \citet{martinache10}, consisting of the sums of translated positive and negative copies of the pupil aperture. Phase offsets outside the support of the pupil have identically zero derivative in the $u,v$ plane as expected. There is a greater magnitude in the derivative at longer baselines compared to the flat maps in \citet{martinache10} because we are rolling together the phase transfer $\mathbf{A}$ matrix and the redundancy matrix $\mathbf{R}$, and short baselines are much more redundant than long ones.

\begin{figure}
    \plotone{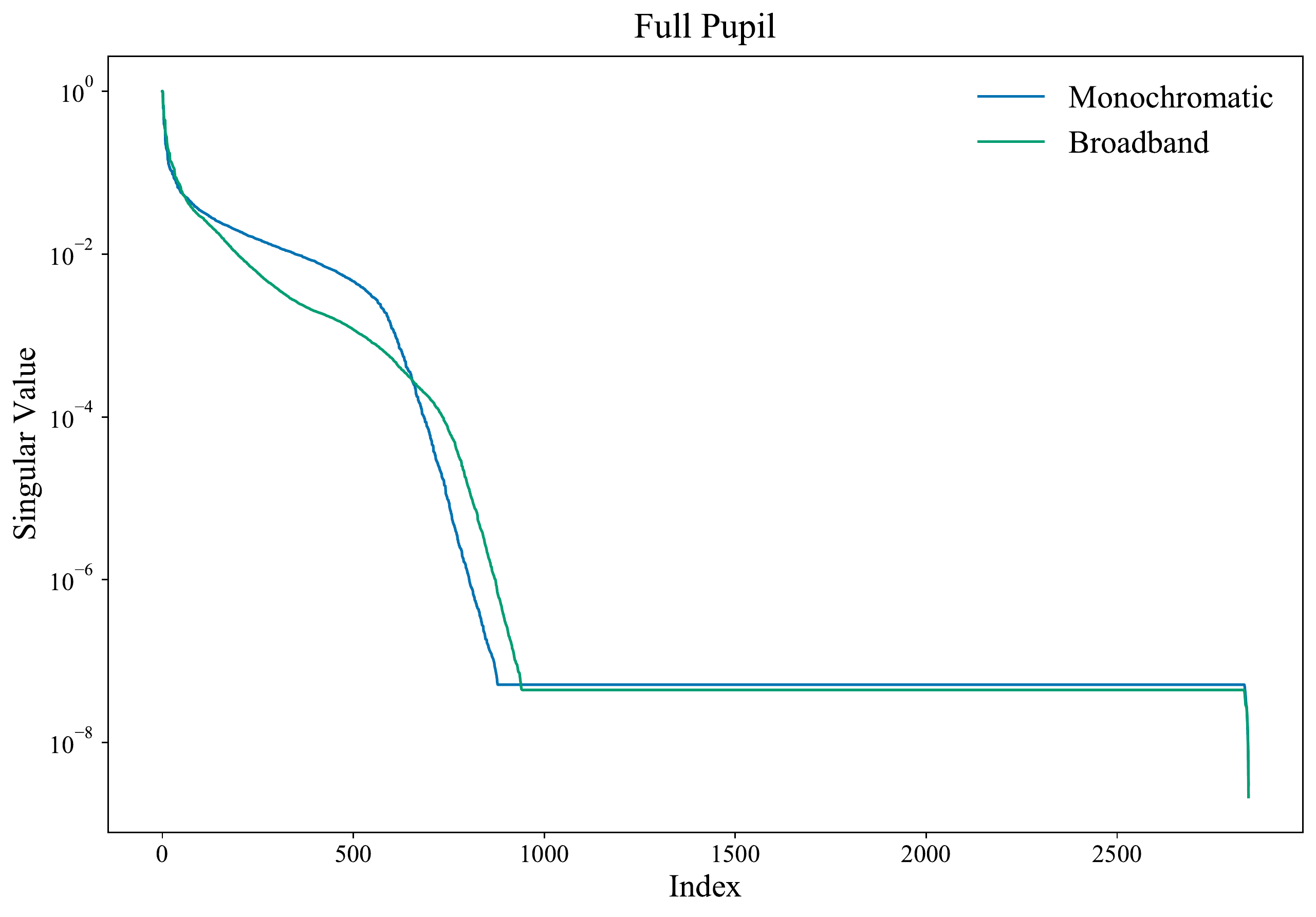}
     \caption{Log-scale plot of the singular values of the phase transfer matrix plotted against index ordered by decreasing singular value for simple monochromatic diffraction (blue), 20\% bandwidth (green). Singular values are normalized to the first singular value. In both cases, there is a sharp cutoff separating significant singular values from a plateau at $\sim 5 \times10^{-8}$ (zero to within machine precision). We take the vectors corresponding to these singular values in the plateau region to be the effective null space of the phase transfer operator, and correspond to kernel phases. \href{https://github.com/benjaminpope/morphine/blob/stable/notebooks/morphine_uv.ipynb}{\color{linkcolor}\faGithub}}
    \label{fig:svd}
\end{figure}

\begin{figure}
\plotone{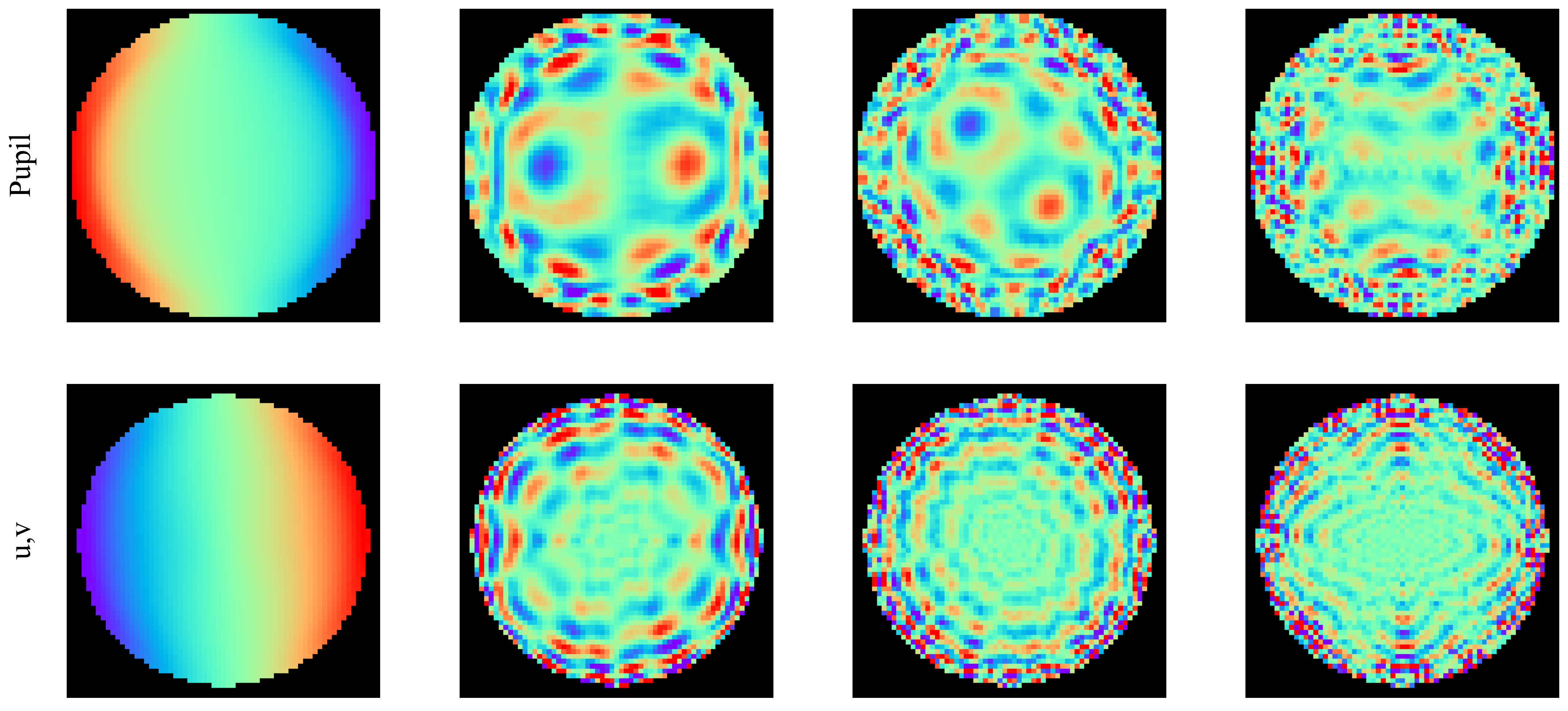}
\caption{Pairs of nonsingular vectors in the pupil and $u,v$ planes from the SVD of the Jacobian in ~\ref{kernel_jacobian}. Some of these orthonormal basis vectors resemble Zernike modes, and all show some ringing structure at the edges of the pupil and OTF support. All eight images are dimensionless and the colour scale is arbitrary. \href{https://github.com/benjaminpope/morphine/blob/stable/notebooks/morphine_uv.ipynb}{\color{linkcolor}\faGithub}
\label{nonsingular_modes}}
\end{figure}

\begin{figure}
\plotone{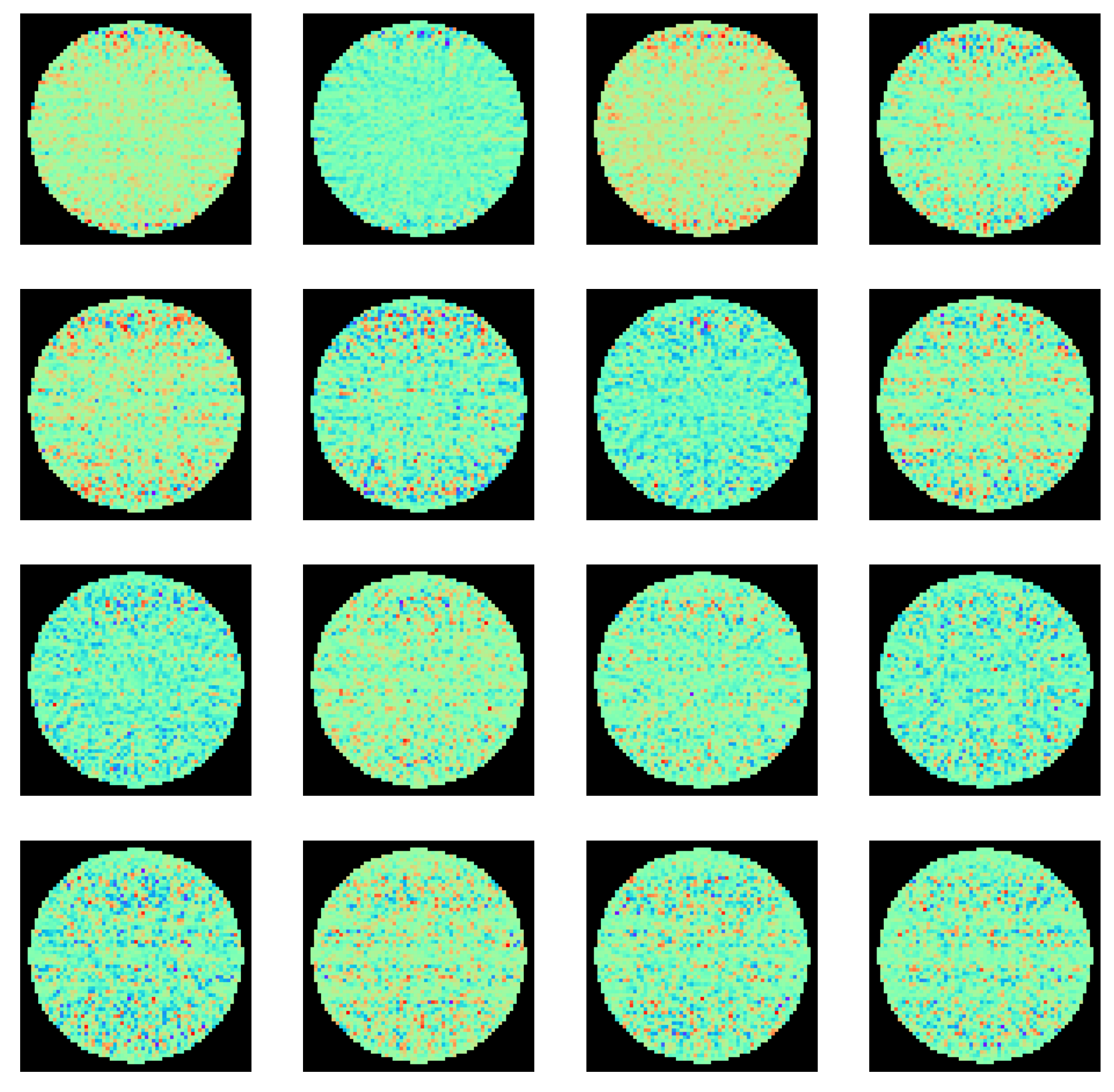}
\caption{Example kernel phase maps - singular vectors in the $u,v$ plane from the SVD of the Jacobian in ~\ref{kernel_jacobian}. These are rows of the $\mathbf{K}$ matrix. There is little apparent structure. All sixteen images are dimensionless and the colour scale is arbitrary. \href{https://github.com/benjaminpope/morphine/blob/stable/notebooks/morphine_uv.ipynb}{\color{linkcolor}\faGithub}
\label{kernel_modes}}
\end{figure}

We calculate the singular values of this matrix (restricted to the support of the pupil) by SVD (Figure~\ref{fig:svd}), finding a sharp cutoff as expected, separating a subspace of kernel phases from phases contaminated by noise from the aberrations. In the nonsingular space, a selection of pairs of corresponding modes in the pupil and $u,v$ plane are displayed in Figure~\ref{nonsingular_modes}, and a selection of kernel phase modes in the $u,v$ plane are shown in Figure~\ref{kernel_modes}. These illustrate how at the edges of the pupil/long baselines there is extra noise. Because of the limited field of view, noise from outside the pupil or optical transfer function (OTF) is convolved inside, and large elements in the transfer matrix can dominate parts of the SVD. For numerical stability we have therefore excluded the very longest baselines in calculating kernel phases, but it would be preferable to use a more robust SVD that is tolerant of outliers. This is not a feature of the autodiff pipeline or \textsc{morphine}, but also of real data. Depending on the field of view of the simulated images, for the same pupil and $u,v$ sampling there can be slight differences in the autodiff-calculated kernel phases because of this effect. The normal analytic kernel phase derivation assumes an infinite field of view, but sees data with a finite field of view in which these window effects are apparent. 

We have also repeated the above calculations for light with a $20\%$ fractional bandwidth, with a uniform spectrum sampled ten times from $1.8\,\mu\text{m}$\,---\,$2.2\,\mu\text{m}$. The results are very similar, except that the Jacobian maps equivalent to Figure~\ref{kernel_jacobian} are slightly blurred. In the shape of the singular value curve between the broadband and monochromatic cases, a qualitatively similar behaviour is found, although the broadband case has slightly fewer kernel phases and a flatter roll-off.

\subsection{Stellar imaging data: $\alpha$ Ophiuchi}
\label{sec:palomar}

We now apply our new kernel phase formalism to real data: the A-type binary star system $\alpha$~Ophiuchi observed with the Palomar~200-Inch telescope using the extreme adaptive optics system PALM-3K and the PHARO camera. These data were first analyzed by \citet{palomar}, successfully recovering the binary and properties of the system. 
These were subsequently improved by \citet{martinache20} who corrected a 2 degree offset in the clocking angle of the real Palomar pupil. Also employing updated `xara' kernel phase code, \citet{martinache20} obtained astrometry of separation $\rho = 123.5 \pm 2.9$~mas, position angle $\theta = 86.5 \pm 0.2$~degrees, and contrast $c = 25.1 \pm 1.1$. The well-analyzed nature of these data make it something of a useful test-piece; we therefore seek to recreate this analysis using our new autodiff methods.

\begin{figure}
\plotone{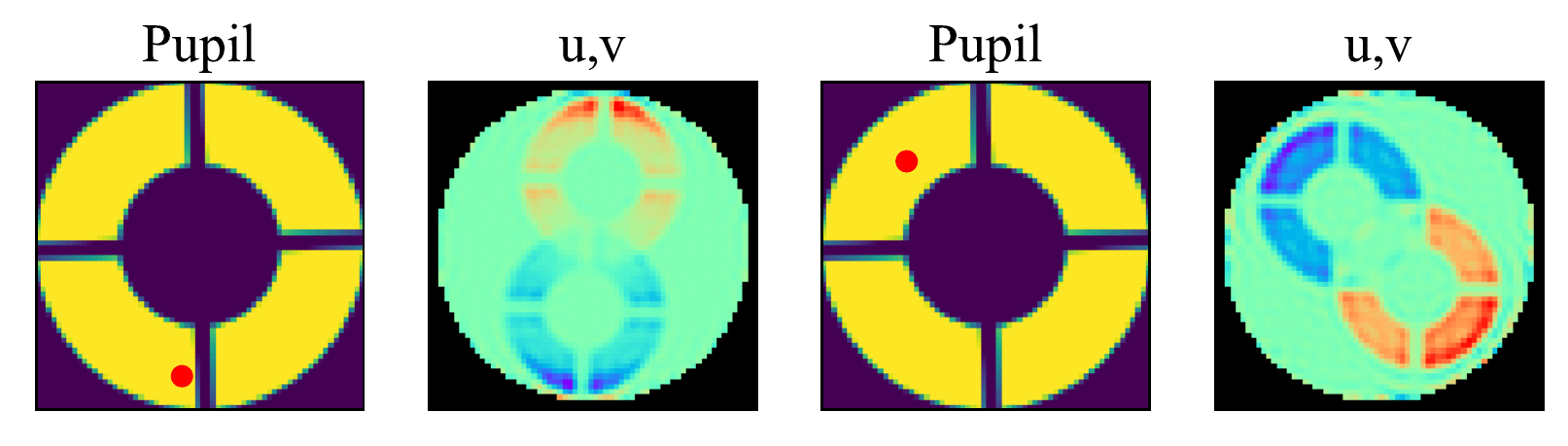}
\caption{A representation of the Jacobian determined by \textsc{jax} for the PHARO pupil in $K$-band. In each pair of images, on the left is the PHARO pupil, with a red dot highlighting a pupil sample; on the right is a map of the derivative in the $u,v$ plane with respect to phase at that pupil sample. These Jacobian vectors are pairs of rows and columns of the $\mathbf{A}$ phase transfer matrix. We see the same maps as in Figure~1 of \citet{martinache10}. \href{https://github.com/benjaminpope/morphine/blob/stable/notebooks/morphine_pharo_xara.ipynb}{\color{linkcolor}\faGithub}\label{pharo_jacobian}}
\end{figure}

We first build a high resolution model of the PHARO pupil using the same code as \citet{martinache20}, and bin it down to a $64\times64$ pixel grid so that pixel values $\in [0,1]$ indicate the fraction of the telescope aperture filling that binned pixel. We then propagate monochromatic 2.145\,$\mu$m light through this onto a $128\times128$-pixel image plane, and use a matrix Fourier transform to map this onto a $u,v$ plane of the same size. We differentiate this using the forwards mode autodiff as above, obtaining the Jacobian shown in Figure~\ref{pharo_jacobian}, and calculate a kernel phase transfer matrix.

\begin{figure}
\plotone{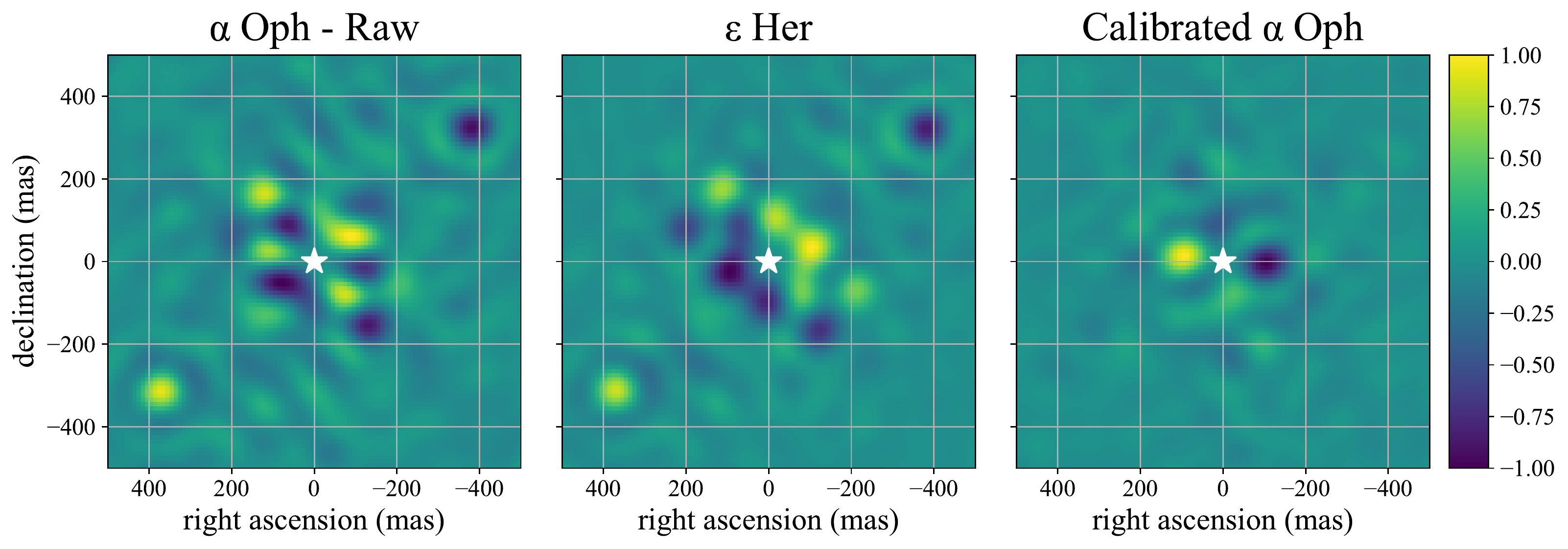}
\caption{Kernel phase colinearity maps showing the normalized dot product of a binary model (pale is a better match) to the kernel phase signals from uncalibrated $\alpha$~Oph data (left), the calibrator $\epsilon$~Her (middle), and calibrated data (right). They closely resemble the maps in Figure~8 of \citet{martinache20}. We can see that in uncalibrated data the filter ghost is clearly visible, but in calibrated data we extract the $\alpha$~Oph binary with no effect from the ghost. \href{https://github.com/benjaminpope/morphine/blob/stable/notebooks/pharo_test.ipynb}{\color{linkcolor}\faGithub} \label{colinearity}}
\end{figure}

We then extract kernel phases from both $\alpha$~Oph and a point source calibrator $\epsilon$~Her using a version of \textsc{xara} modified for compatibility with \textsc{morphine}. Following \citet{martinache20}, we choose as our observables the median kernel phases across all 100 exposures for each, and the standard error of the mean as our base uncertainty on each. To calibrate the kernel phases, we simply subtract those of the calibrator from $\alpha$~Oph and add their uncertainties in quadrature. To see the effect of the known optical `ghost' (due to an unwanted reflection from the neutral density filter used to observe bright stars such as these), we calculate kernel phase `colinearity' maps as in \citet{martinache20}, shown in Figure~\ref{colinearity}. In these maps we can see that the filter ghost shows up strongly as a false binary in the uncalibrated kernel phase maps, but calibrated kernel phases easily remove this and reveal the tight $\alpha$~Oph binary. 

In \citet{martinache20} a large error term was added in quadrature in order to account for additional noise in the data unexplained by the diversity over exposures and uncalibrated by the kernel phase model. Using nonlinear least squares to fit the calibrated data, we also find that the best-fitting binary model has a high $\chi^2$. We therefore follow \citet{martinache20} and add a dimensionless kernel phase quantity of 0.0184 in quadrature to the uncertainties to make the reduced $\chi^2 = 1.0$.

\begin{figure}
\plotone{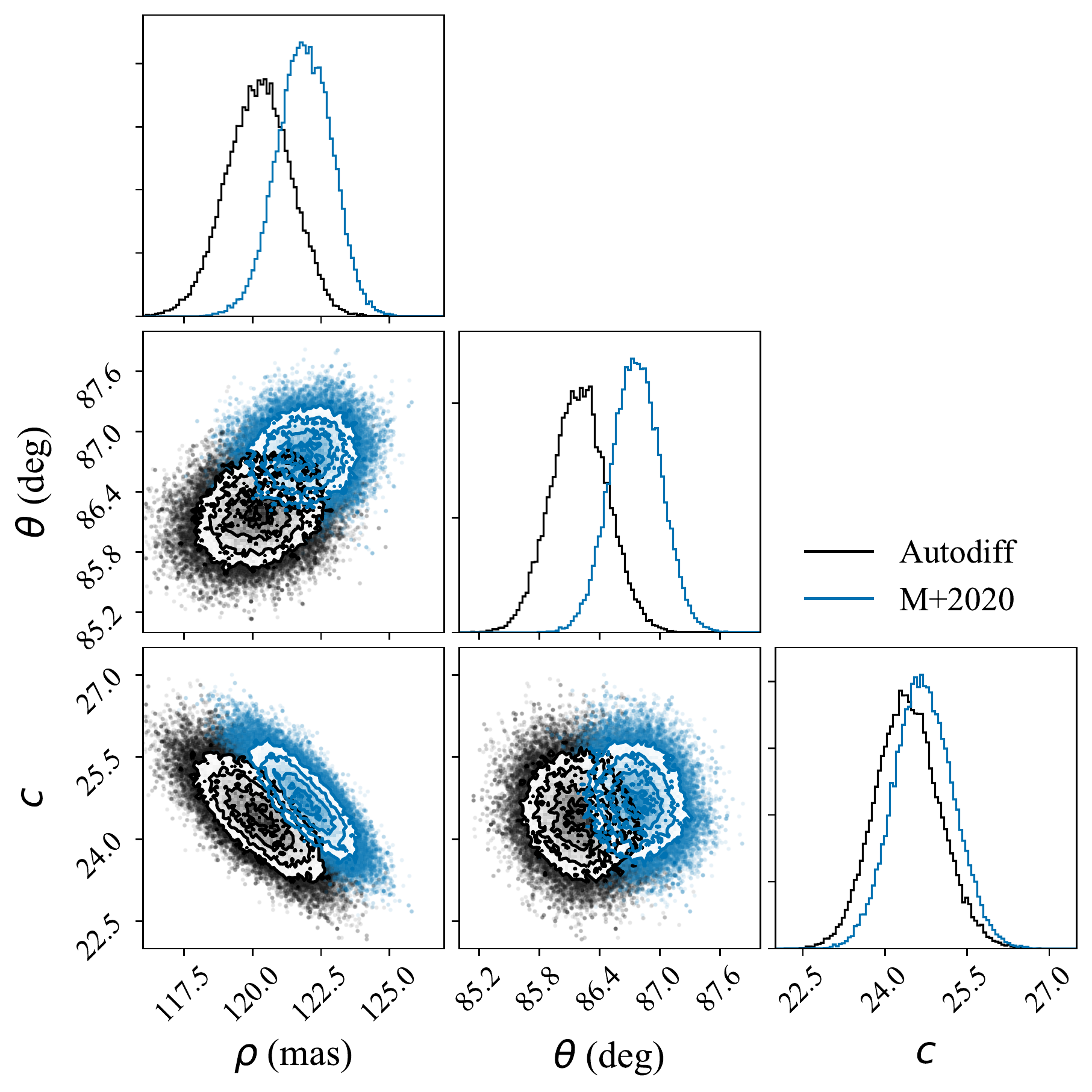}
\caption{Posterior distributions for the binary parameters of $\alpha$~Ophiuchi as determined by the autodiff pipeline (black) and the \citet{martinache20} implementation (blue). \href{https://github.com/benjaminpope/morphine/blob/stable/notebooks/frantz_test.ipynb}{\color{linkcolor}\faGithub} \label{comparison_posterior}}
\end{figure}

We then use Markov Chain Monte Carlo \citep{metropolis53} to sample from the posteriors assuming Gaussian likelihoods, using the \texttt{emcee} affine-invariant ensemble sampler \citep{emcee}. We obtain binary parameters of separation $\rho = 120.2 \pm 1.2$~mas, position angle $\theta = 86.2 \pm 0.3$~degrees, and contrast $c = 24.4 \pm 0.6$: all well within 1$\sigma$ of the \citet{martinache20} figures. We likewise use \texttt{emcee} to sample from the likelihood using the \citet{martinache20} code. Posteriors for both inferences are shown in Figure~\ref{comparison_posterior}.

\subsection{Simulated binary star data}

\begin{figure}
\plotone{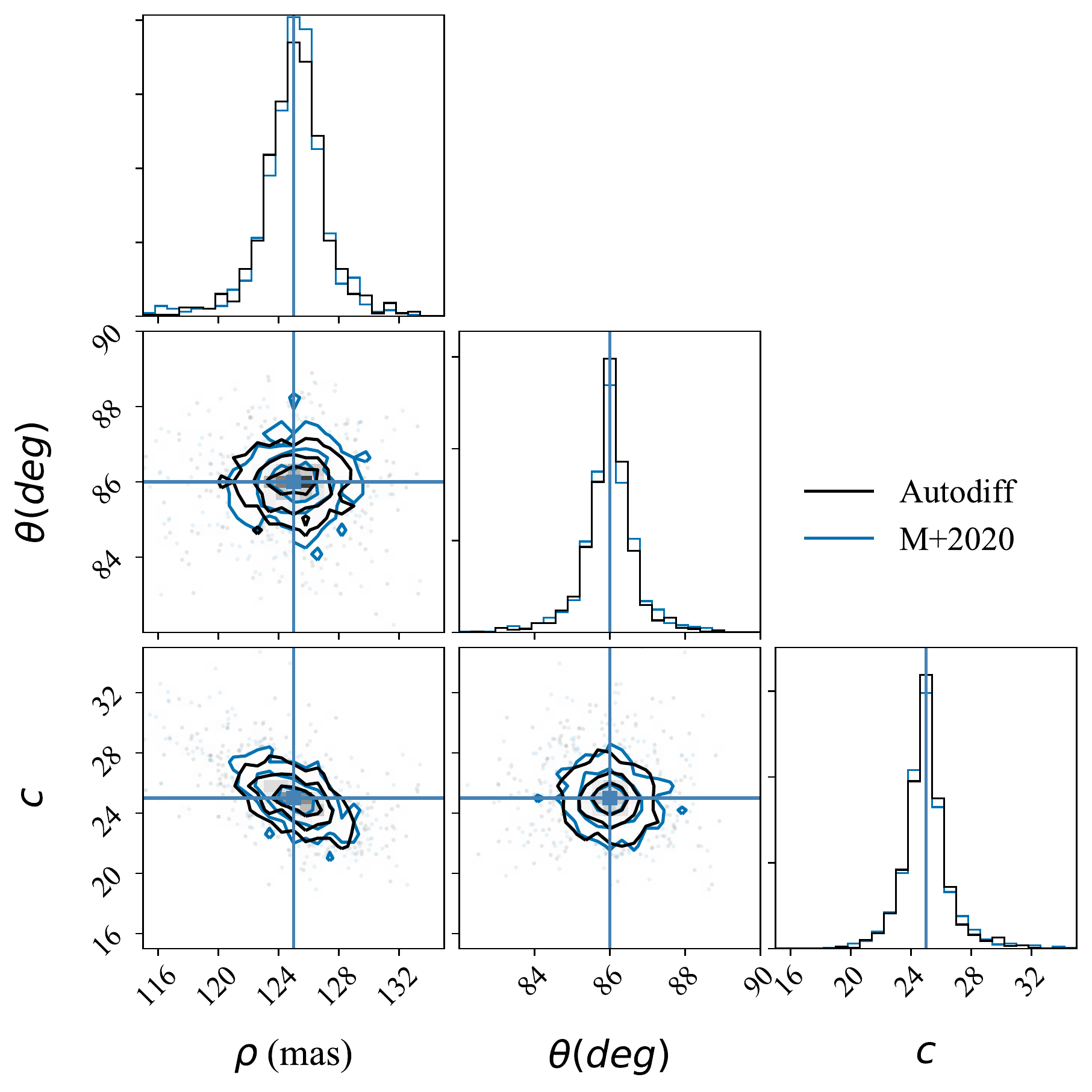}
\caption{Comparison of \citet[][blue linetype]{martinache20} and autodiff (black) kernel phase pipelines on simulated data with true parameters $\rho = 125$~mas, $\theta = 86^\circ$, $c=25$. We generate a cube of 1000 mock datasets using a 64-pix pupil model and Zernike aberrations as described in Section~\ref{sec:palomar}. The corner plot displays the parameters retrieved using a simple least squares fit applied to each individual frame. The methods have very similar distributions centred tightly on the true value. \href{https://github.com/benjaminpope/morphine/blob/stable/notebooks/frantz_test_sim_cube.ipynb}{\color{linkcolor}\faGithub}
\label{comparison_posterior_sim_cube}}
\end{figure}

In order to investigate the effect on both pipelines of wavefront errors leaking in, we generated a cube of 1000 PSFs at a wavelength of $2.145\,\mu$m with 20 Zernike modes of aberrations, this time with 20\,nm standard deviation, for a total peak-to-peak wavefront amplitude of $295\pm 100$\,nm. We then retrieved the system parameters with \texttt{emcee} for the cube
frame-by-frame using weighted least squares (displayed in Figure~\ref{comparison_posterior_sim_cube}). For a simulated binary with fiducial $\rho = 125$\,mas, $\theta=86^\circ$, and $c = 25$, there is no sign of the large systematic difference between the methods as seen in real data. Both methods, whether frame by frame or on average, retrieve the true parameters with similar accuracy and precision. 

\begin{figure}
    \plotone{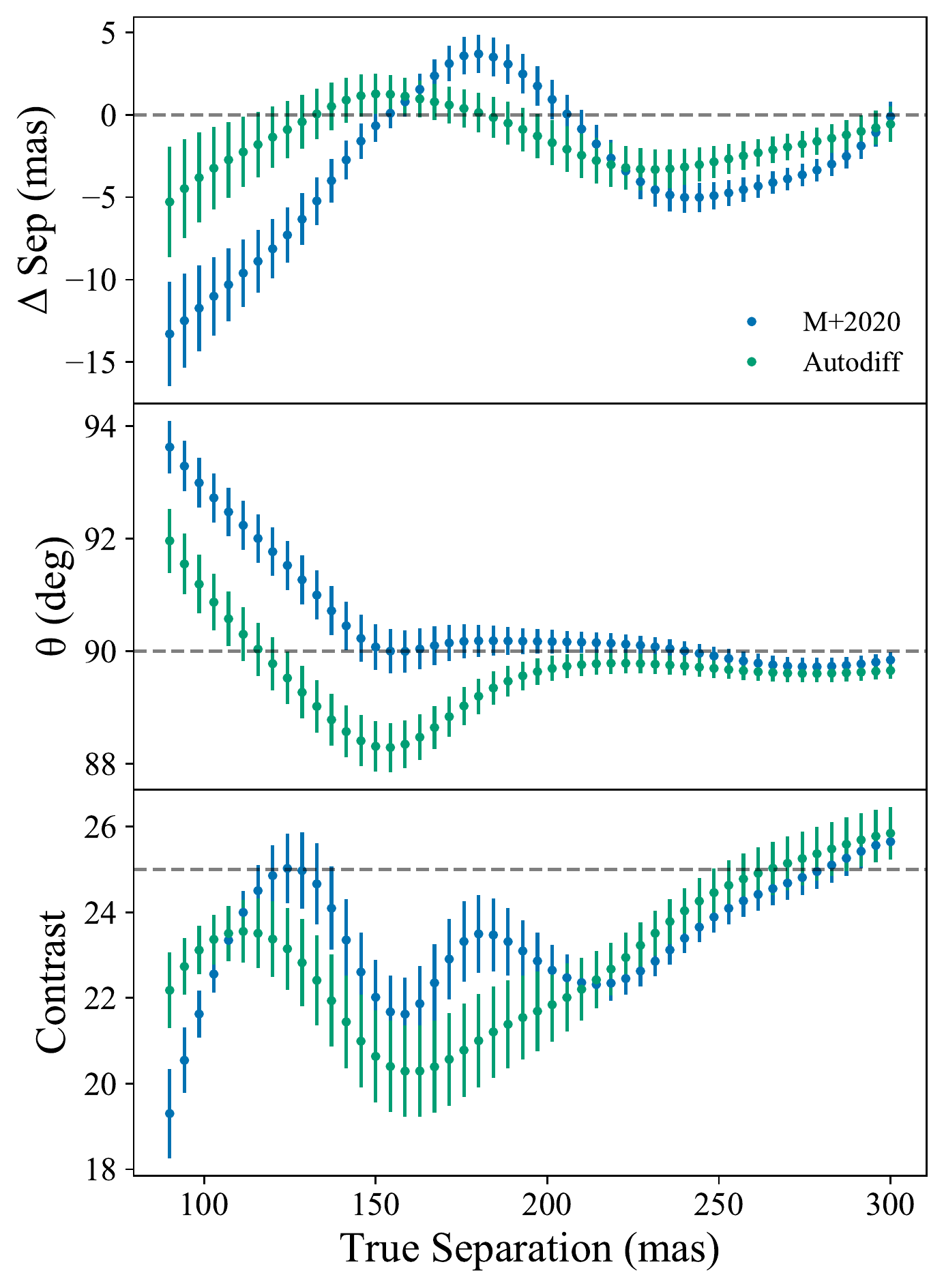}
     \caption{Injection recovery results for PHARO using both the \citet{martinache20} and autodiff kernel phase models. From top to bottom we show the error in the recovered separation, the recovered position angle, and the recovered contrast as a function of injected separation, with $1\sigma$ uncertainty errorbars. Both methods show similar performance, with small systematic errors of comparable size to their statistical uncertainties. \href{https://github.com/benjaminpope/morphine/blob/stable/notebooks/frantz_test.ipynb}{\color{linkcolor}\faGithub}}
    \label{injection_recovery}
\end{figure}

In order to establish the accuracy of both pipelines, we conducted an injection test. We took the first frame of the $\epsilon$~Her datacube as our `data' and use a subpixel Fourier shift to create a mock binary at a position angle of $90^\circ$, contrast 25, and a range of separations from 90-300~mas. We then used least squares to fit these data using both models (initialized at the true values), determined in each case an error to add in quadrature to make reduced $\chi^2=1.0$, and re-ran the least squares fit with new data, saving the best-fit parameters and uncertainties taken from the square root of the diagonal elements of the inverse covariance matrix. The results are shown in Figure~\ref{injection_recovery}. Both methods perform comparably well, achieving results reasonably close to the injected values. Systematics in all three parameters, of amplitude comparable to the $1\sigma$ statistical uncertainties, vary smoothly and similarly as a function of input separation. The autodiff pipeline performs somewhat better at smaller separations in all three parameters. This indicates that both kernel phase pipelines do a good overall job at binary imaging with PHARO, but that as noted by \citet{martinache20} not all sources of uncertainty are well-understood or calibrated.

\subsection{Complex optical systems: nullers and coronagraphs}
\label{sec:coronagraph}

It is desirable to find a priori self-calibration schemes for high-angular-resolution, high-contrast imaging systems more generally such as nulling interferometers and more optically complex coronagraphs intended to suppress light in a region around a target star. One particularly interesting extension of the kernel phase idea is the `kernel nuller' concept \citep{martinache18}, in which the aberration transfer matrix and kernel operator idea is applied to the output of a nulling interferometer. 

For arbitrary systems, to generate our Taylor expansion, we wish to be able to take the Jacobian of arbitrary features of the final speckle pattern with respect to phases in the input wavefront or in intermediate planes. This could then be used in postprocessing to generate `kernel speckles' resistant to noise, or in wavefront control for adaptive optics, or for digging `dark holes' to search for high contrast companions \citep{malbet95}.

Arbitrary-order Taylor series expansions for a PSF with respect to small phase perturbations can be derived analytically for a telescope with an arbitrary pupil brought to a single focus \citep{bloemhof01,anand02,perrin03}. This expansion breaks down for more complex optical systems. The PSF of a coronagraphic imager far from the occulted region is very similar to that of the primary PSF of the telescope, but close to the inner working angle it is significantly affected by the occultation and especially strongly by low-order wavefront errors. This is likewise an issue where detector nonlinearity and saturation significantly distort the PSF. While the analytic series expansion is no longer applicable in these regimes, using matrix Fourier transforms and skipping saturated pixels or pixels inside the coronagraphic inner working angle has shown promise for pushing kernel phase beyond its conventional limitations \citep{laugier19b}, as have generalizations that exploit angular differential imaging for a further level of calibration \citep{laugier20}.

Here we will apply the formalism demonstrated above for simple diffraction to a new case: coronagraphy. We will find that the identical approach generates an analogous modal decomposition and kernel observables.

We consider a very simple Lyot coronagraph \citep{lyot30}. We choose a 1\,$\mu$m wavelength, and 1\,m diameter pupil sampled on a $96\times96$ grid. The central $4 \lambda/D$ region of the resulting PSF is blocked out with an on-axis focal plane stop (occulter), and the light is then propagated to a second pupil plane using an FFT rather than a matrix FT, following standard \textsc{poppy} practice. This imposes tougher memory constraints, and in future may be replaced with an MFT. A Lyot stop (an iris mask undersized 10\% relative to the input pupil) is then imposed, and the light propagated to a final focal plane sampled on a 50\,mas pixel scale grid with a 4~arcsec field of view. Aside from the coarse gridding, this is intentionally identical to one of the standard test and verification examples supplied with \textsc{poppy}\footnote{\href{https://github.com/mperrin/poppy/blob/master/notebooks/MatrixFTCoronagraph_demo.ipynb}{github.com/mperrin/poppy/blob/master/notebooks/MatrixFTCoronagraph\_demo.ipynb}}. The PSF produced has a dark coronagraphic hole in the middle, and beyond an inner working angle around this hole has a pattern of diffraction rings.

\begin{figure}
\plotone{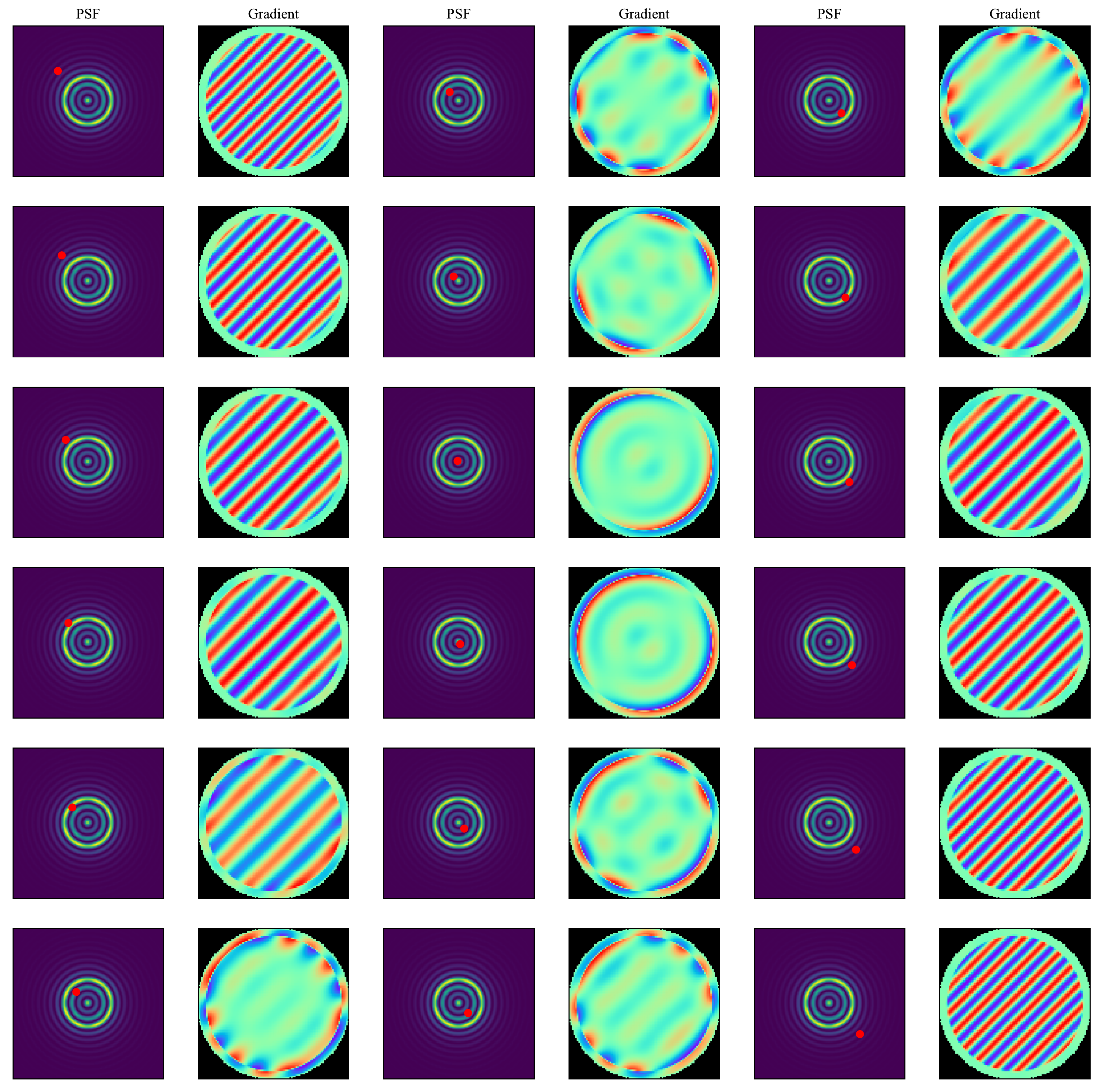}
\caption{A representation of the Jacobian determined by \textsc{jax} for a simple Lyot coronagraph. In each pair of images, on the left is the PSF, with a red dot highlighting the selected pixel; on the right is a map of the corresponding pixel's derivative with respect to phase over the pupil, the pattern of pupil phases that linearly affect that speckle. Speckles far from the central dark hole are generated by sinusoids across the pupil, while speckles near the inner working angle correspond to more complicated patterns.  \href{https://github.com/benjaminpope/morphine/blob/stable/notebooks/morphine_coronagraph.ipynb}{\color{linkcolor}\faGithub}
\label{speckle_jacobian}}
\end{figure}

Again using \textsc{jax}, we calculate the Jacobian of this PSF with respect to the input wavefront, evaluated at uniform-zero phase, which again takes a few minutes on a laptop. The previous approach has been to directly propagate through the end to end simulation for a grid of small perturbations in the input plane \citep{falco}, whereas this is now taken care of by autodiff. The results are displayed in Figure~\ref{speckle_jacobian} in a similar format to Figure~\ref{kernel_jacobian}. Pixels far out in the PSF correspond to sinusoidal modulation of the input wavefront as expected, and this is symmetric about the origin. On the other hand, pixels near the inner working angle where the effects of the coronagraph are evident correspond to more complex modes in the pupil, also containing contributions from the ring obscured by the Lyot stop when the pupil is re-imaged. 

\begin{figure}
    \plotone{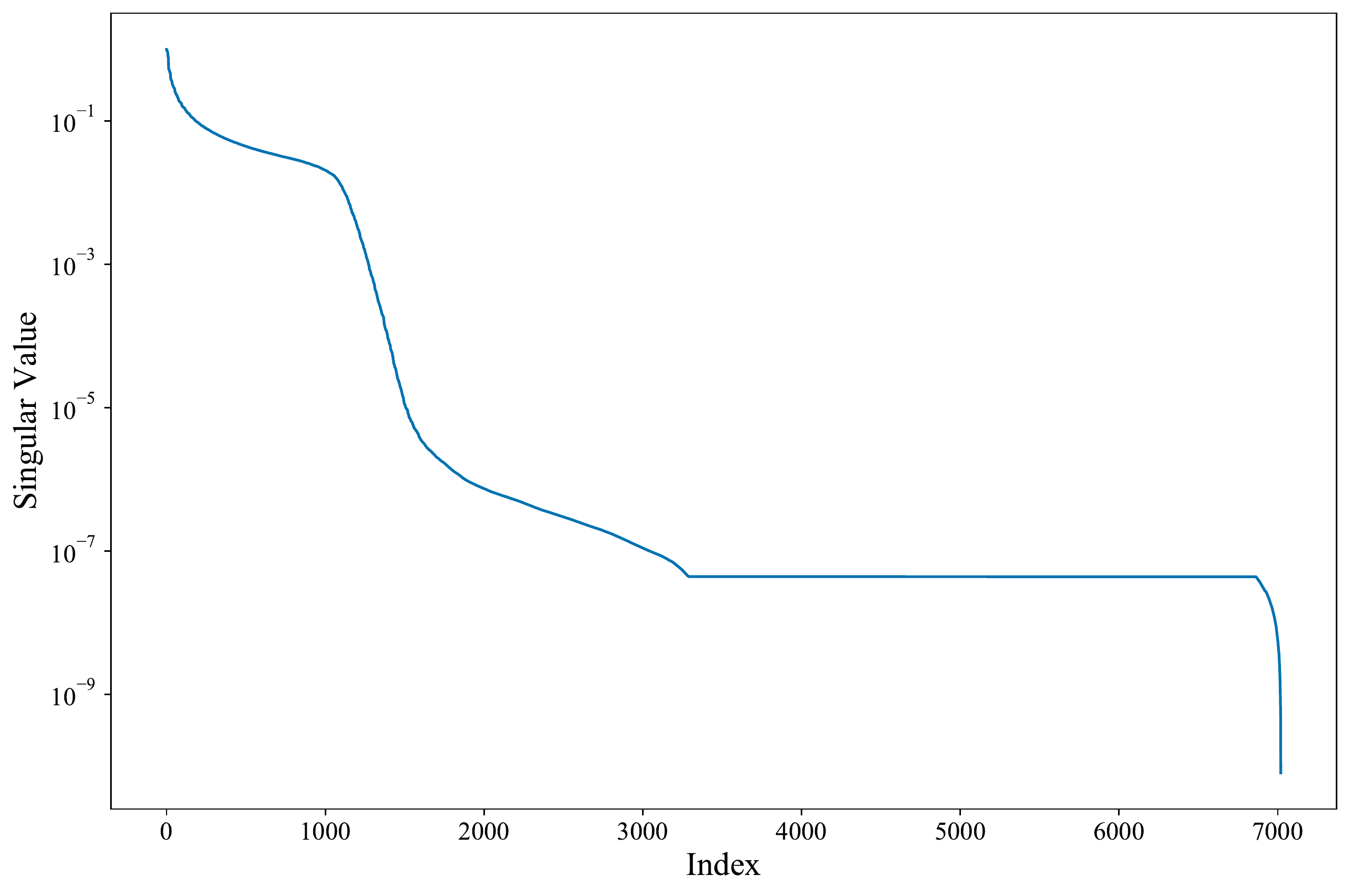}
     \caption{Log-scale plot of the singular values of the phase transfer matrix versus index for the coronagraph simulation in Section~\ref{sec:coronagraph}. Singular values are normalized to the first singular value. \href{https://github.com/benjaminpope/morphine/blob/stable/notebooks/morphine_coronagraph.ipynb}{\color{linkcolor}\faGithub}}
    \label{fig:svd_coronagraph}
\end{figure}

\begin{figure}
\plotone{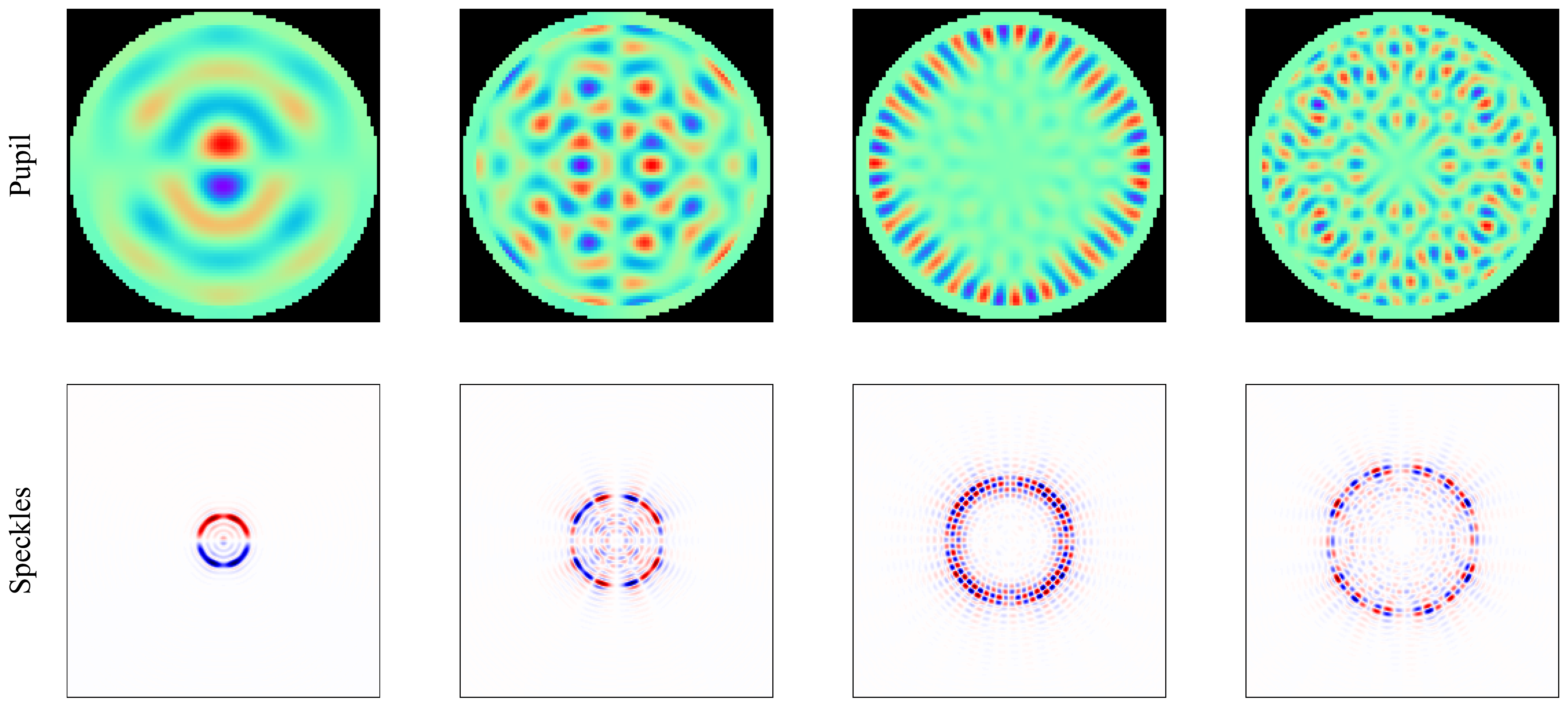}
\caption{Pairs of nonsingular vectors in the pupil and image planes from the SVD of the coronagraph Jacobian in~\ref{speckle_jacobian}. \href{https://github.com/benjaminpope/morphine/blob/stable/notebooks/morphine_coronagraph.ipynb}{\color{linkcolor}\faGithub}
\label{nonsingular_corona}} 
\end{figure}

Calculating the SVD of this Jacobian (again restricted to the support of the primary pupil) we see a decline to very small singular values. The ordered singular value curve is displayed in Figure~\ref{fig:svd_coronagraph}. The nonsingular modes form a basis, like the Zernike basis, that maps between aberration patterns in the pupil and the speckle modes they generate. Some pairs of nonsingular modes in the pupil and image plane are shown in Figure~\ref{nonsingular_corona}

\begin{figure}
\plotone{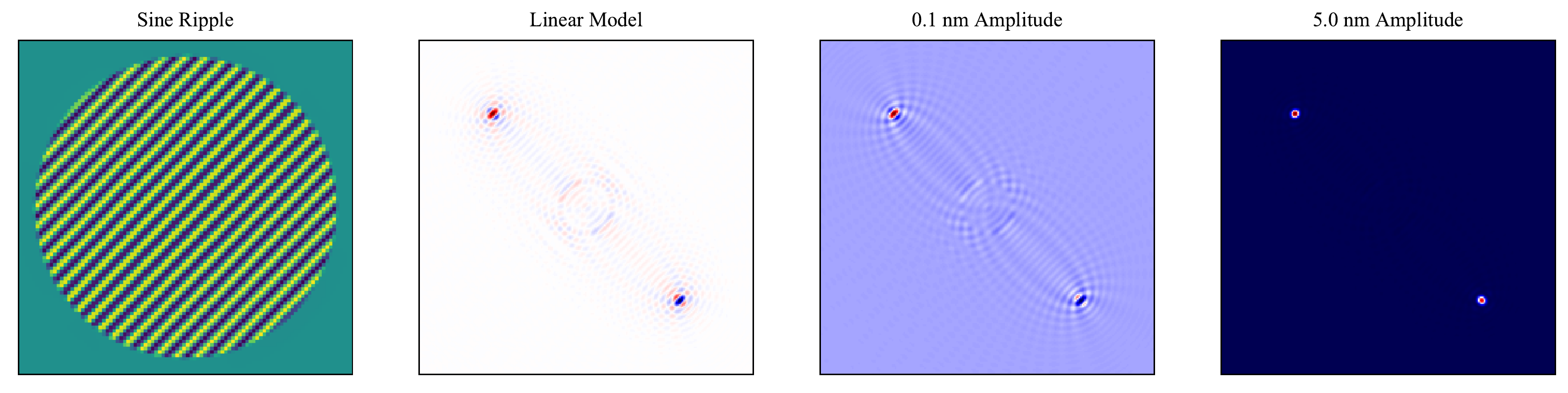}
\caption{The effect of a sinusoidal phase ripple on the speckle pattern of the coronagraph in Section~\ref{sec:coronagraph}. Left: sinusoidal phase pattern. Centre left: linear transfer map applied to the sinusoidal phase pattern, showing alternating positive and negative speckles on either side of the centre of the PSF. Centre right: the difference between a full optical simulation with an 0.1~nm amplitude phase ripple with respect to an unaberrated PSF, clearly close to the linear case. Right: the same difference simulation but with a 5~nm amplitude ripple, showing positive speckles on both sides of the detector - the quadratic term. \href{https://github.com/benjaminpope/morphine/blob/stable/notebooks/morphine_coronagraph.ipynb}{\color{linkcolor}\faGithub}
\label{corona_nonlinearity}}
\end{figure}

For the coronagraph under consideration, the regime over which the linear approximation applies may be strictly limited. To illustrate this, we take one of the sinusoidal ripples which are the columns of the Jacobian in Figure~\ref{speckle_jacobian}, propagate this through a full optical simulation, and take the difference with respect to an unaberrated PSF. In Figure~\ref{corona_nonlinearity}, we see the effects of a sinusoidal phase ripple in the nearly-exact linear regime with a 0.1~nm amplitude, and with a small but non-negligible 5.0~nm amplitude. In the linear regime (where the phase ripple is very small) we see that the speckle is positive on one side of the PSF and negative on the other, corresponding closely with expected behavior from the Jacobian. On the other hand, increasing the amplitude of the phase ripple to just 5 nanometres is enough to alter the outcome completely. The figure indicates a switch in behavior so that the speckles are now both positive, indicating the quadratic term in the Taylor series has become dominant. This quadratic term involves a rank-3 tensor, of the second partial derivatives of each pixel with respect to the wavefront -- and there is no equivalent of a kernel operator for tensors of this dimension. The use of kernel phases as robust observables will therefore fail in this case, though we suggest that future work may search for a higher-dimensional kernel manifold. Rigorously the transition between linear and quadratic regimes occurs when the gradient of the contrast with respect to a given mode is not constant anymore. For small wavefronts, e.g $\epsilon \rightarrow 0$,  the constant gradient is written as $ b$, e.g $ \frac{\partial C } {\partial \epsilon} \rightarrow b$. Then the transition is defined by  $ |\frac{1}{b} \frac{\partial C} {\partial \epsilon}  - 1| = 1$. Numerically, this relationship can be inverted to find the wavefront at the transition by using the autodiff tools in this paper. In practice it occurs when the contrast is three to five time greater than the coronagraph's raw contrast (without any wavefront errors). 

We have therefore shown that under the same linear assumptions previously applied to yield kernel phases, analogous self-calibrating observables also exist for propagation through more complex optical systems such as coronagraphs. However for the example configuration explored, the rapid onset of non-linear response indicates they may be of limited practical applicability to coronagraphic observations except in the extremely high-wavefront-quality regime. Sub-nanometre wavefront precision may nevertheless be achieved in future large space telescopes with high-order deformable mirrors.

\subsection{Zernike Basis}
\label{zernike}

Taking the Jacobian of one large pixel grid by another is expensive in terms of memory; it is therefore desirable to bring down the dimensionality of either the input or target space. One way to do this is to use a Zernike basis to represent the wavefront \citep{zernike34}. One advantage of this is that we can examine only up to a certain order in Zernike polynomials if we believe that wavefront errors are negligible below a certain spatial scale. We can then differentiate just with respect to Zernike coefficients, which is much more computationally efficient and allows us to run simulations on grids that are many times larger than otherwise. Similar bases (eg the `hexike' basis) can be constructed for arbitrary non-circular telescope apertures.

We have re-run the simple diffraction simulation from Section~\ref{sec:simple} with 200 Zernike modes on a $256\times256$ grid, and the coronagraph from Section~\ref{sec:coronagraph} on a $128\times128$ grid using 300 Zernike modes. The calculations now take around 4 seconds, rather than many minutes, with peak memory use around 1.5 and 3~GB rather than 7~and 9~GB respectively. 

The Jacobian derived from this basis for simple diffraction is an excellent approximation to the full calculation on a grid. Given that in problems such as HST data analysis we believe that PSF variations are dominated by low-order modes \citep{lallo06}, this allows us to calculate a larger kernel space with a more accurate pupil model than has previously been possible. 

While it is a huge advantage in simulation speed and fidelity, this basis does not accurately represent the high-spatial-frequency sine waves corresponding to speckles far from the core of the PSF. Nevertheless, it does an adequate job representing the effects of low-order wavefront error, and this may be sufficient for many purposes, for example in optimization of optical design.

\section{Discussion}
\label{sec:discussion}

A phase transfer matrix constructed analytically with a discrete pupil model contains no a priori information about the pixel sampling, binning, or windowing. This means that the implicit convolutions in the $u,v$ plane, and therefore the associated correlation between adjacent baselines, is not taken into account.  Therefore the normally-orthonormal kernel phases (calculated analytically for an infinite field of view and fine sampling) are no longer fully linearly independent. For example \citet{martinache20} notes that in the analysis of \citet{palomar}, windowing to avoid the ghost introduced by a neutral density filter cuts the number of pixels to less than the number of kernel phases, so that information is being lost. \citet{martinache20} also demonstrates the utility of a detailed model of pupil (mis-)alignment to improve the extraction of both kernel phase and visibility information.

The non-independence of kernel phases has been treated statistically by \citet{ireland13}, who uses principal component analysis applied to an ensemble of kernel phase observations to extract statistically-independent kernel phases. When the advantage of kernel phase is knowing that some combinations of phases have high signal-to-noise \textit{a priori}, it would be preferable to avoid this situation with a better kernel phase construction rather than calibration.

A further issue is that the $u,v$-plane calculation of earlier kernel phase work suffers from discretization noise from interpolating the FFT. This can be ameliorated by using a matrix Fourier transform adapted to the wavelength and exact pupil model \citep{martinache_habilitation}.

By calculating kernel phases from differentiating an end-to-end optical simulation, even for trivial apertures, the exact pixel grid, including binning, windowing, and dead pixels can be included. As the matrix DFT is already used to calculate the FT, there is no sampling noise to separately incorporate, and all kernel phases are automatically orthonormal.

To avoid self-subtraction, we expect that the best approach to data analysis with kernel phases using these methods will be forwards-modelling the data through the optical system and kernel phase filter along the lines described by \citet{pueyo16} or \citet{martinache20}. These can then be augmented by diversity-based methods along the lines described by \citet{ireland13} to achieve even higher levels of calibration. We have not attempted to use these methods on real data in this paper.


Because of its \textsc{NumPy}-like API, we have used \textsc{jax} to power the automatic differentiation in \textsc{morphine}. We have not extensively optimized this code for fast and memory-efficient autodiff, nor have we benchmarked it against competing frameworks such as \textsc{TensorFlow}. We expect that integration with \textsc{NumPy} and similarity to the existing standard \textsc{poppy} will make \textsc{morphine} more readily useful in astronomy, but for integration with neural networks it may be more convenient to use a different autodiff package. 

\section{Open Science}
\label{sec:open}

In the interests of open science, we have made available the Python package \textsc{morphine}, together with Jupyter notebooks used to generate the figures in this paper, under a BSD 3-clause open source license at \href{https://github.com/benjaminpope/morphine}{github.com/benjaminpope/morphine}. We encourage and welcome other scientists to replicate, apply, and extend our work.

\section{Conclusions and Future Work}
\label{sec:conclusions}

We have shown that automatic differentiation software offers significant benefits for simulating astronomical optics. Using a simple model of diffraction, we can recover the idea of kernel phases, and using real Palomar data we obtain similar astrometric performance to the analytically-derived standard approach. Crucially, this can then be extended to arbitrary systems for which analytic self-calibration may not be possible. We have demonstrated this for a simple Lyot coronagraph, finding a modal basis and kernel observables analogous to kernel phases. However this is unfortunately of limited practical utility, due to the strong nonlinearities in the Lyot coronagraph problem studied. Nevertheless, alternate coronagraphs and imaging systems such as vortex coronagraphs \citep{foo05}, phase-induced amplitude apodization \citep[`PIAA'][]{guyon03} or nulling interferometers \citep{bracewell78} may derive greater benefit from analysis with this approach. As well as this, second- or higher-order Taylor expansion of the aberrations may reveal higher-order kernel phases even in the Lyot case (Xin et al, in prep.).

In this paper we have shown that autodiff allows us to straightforwardly account for broadband effects as a linear combination of diffraction simulations, which are also differentiable. Likewise, linear mixing of polarization states via Mueller matrices could be included in this framework. This may be important for high angular resolution differential-polarimetry instruments such as SPHERE/ZIMPOL \citep{zimpol} on the VLT, VAMPIRES on the Subaru SCExAO \citep{vampires}, or the GPI polarimeter \citep{gpipol}, where polarization measurements are intrinsic to the goal of the experiment, or to issues with polarization leakage affecting closure phases in holographic aperture masking experiments \citep[e.g.][]{doelman18}. 

In high contrast imaging with adaptive optics, a major issue of residual aberration comes from differing optical paths: both those between the deformable mirror and the wavefront sensor, and the deformable mirror to the science camera. Non-common-path aberrations in this part of the optical system cannot be sensed and corrected inside the AO loop alone. Slowly-evolving non-common path errors cause quasi-static speckles that are one of the key sources of noise limiting the sensitivity of high-contrast instruments to exoplanets. In systems with an asymmetric pupil, the kernel phase formalism can also be used for focal-plane wavefront sensing \citep{martinache13}. This has the advantage of avoiding non-common path errors between the wavefront sensor and the science instrument. 
This can be made even more effective with an integral field spectrograph, as spectrally-dispersed PSFs encode a hologram of the input wavefront \citep{martinache16}. The asymmetric pupil linear phase transfer approach has been used for focal-plane wavefront sensing in the lab \citep[e.g.][]{pope14,swift} and on Subaru SCExAO \citep{martinache16b}. Extending the domain of applicability of the linear phase transfer paradigm to coronagraphs using autodiff may allow for new approaches to focal plane wavefront sensing. 

The approach of \citet{sitzmann2018} to end-to-end optimization of an optical system has clear relevance to optical astronomy. Going beyond using gradients for phase retrieval \citep{jurling14}, we can envision gradient-based \textit{phase design} for coronagraphs,  for diffractive-pupil telescopes, or for starshades. For example, it may be useful in rapidly finding deformable mirror settings to generate a ``dark hole'' in high-contrast imaging \citep[e.g.][]{malbet95,currie20}. As noted above, autodiff has already been applied extensively in photonics; there may be many useful applications in astrophotonics, comining astronomical imaging optics with photonic devices such as beam combiners, photonic lanterns, and single-mode spectrographs \citep{jbh17,minardi20}. It may also be useful for diffractive-pupil design: for example, the \textsc{Toliman} space telescope concept \citep{tuthill18,bendek18} aims to use a pupil-plane phase mask to achieve high precision relative astrometry of $\alpha$~Centauri AB to search for exoplanets. Work so far in designing the phase mask has been in optimizing parametrized pupils with respect to the radially-weighted gradient energy of the PSF as a proxy for astrometric precision. It will be in principle possible to calculate the Fisher information for the astrometric precision, and to use this as an objective function for optimizing a non-parametric \textsc{Toliman} pupil. By differentiating with respect to phase in intermediate planes, \textsc{morphine}-like methods will also be useful in intermediate-plane wavefront control and mask design.

The generalization of kernel phase and optical gradient design permitted by autodiff also allows us to extend this work to the near-field (Fresnel) propagation regime. This may lead to improvements in modelling of components (such as deformable mirrors) that might occur in planes intermediate between pupil and focus, or simply more accurate kernel phase models of instruments such as the \textit{Hubble Space Telescope}. This may be most relevant outside of astronomy: for example, Fresnel coherent diffractive imaging \citep{williams2006} is a popular microscopy technique, in which gradient-based advances in phase retrieval have been applied \citep{Dueaay3700}, and kernel phase and optimization may be valuable. 

\section*{Acknowledgements} 

We would like to thank Anand Sivaramakrishnan, Marshall Perrin, Will Farr, David Hogg, Louis Desdoigts, Alison Wong, and Nour Skaf for their very helpful comments. 

This work was performed in part under contract with the Jet Propulsion Laboratory (JPL) funded by NASA through the Sagan Fellowship Program executed by the NASA Exoplanet Science Institute. 

This research made use of NASA's Astrophysics Data System.

BJSP acknowledges being on the traditional territory of the Lenape Nations and recognizes that Manhattan continues to be the home to many Algonkian peoples. We give blessings and thanks to the Lenape people and Lenape Nations in recognition that we are carrying out this work on their indigenous homelands. We acknowledge and pay respect to the Gadigal people of the Eora Nation. It is upon their ancestral lands that the University of Sydney is built.

\software{This research made use of \textsc{jax} \citep{jax}; \textsc{poppy}, an open-source optical propagation Python package originally developed for the James Webb Space Telescope project \citep{poppy}; the \textsc{IPython} package \citep{ipython}; \textsc{NumPy} \citep{numpy}; \textsc{matplotlib} \citep{matplotlib}; \textsc{SciPy} \citep{scipy}; \texttt{emcee} \citep{emcee} and \texttt{corner.py} \citep{corner}; and Astropy, a community-developed core Python package for astronomy \citep{astropy}.}



\bibliography{ms}

\begin{thebibliography}{}
\expandafter\ifx\csname natexlab\endcsname\relax\def\natexlab#1{#1}\fi
\providecommand{\url}[1]{\href{#1}{#1}}
\providecommand{\dodoi}[1]{doi:~\href{http://doi.org/#1}{\nolinkurl{#1}}}
\providecommand{\doeprint}[1]{\href{http://ascl.net/#1}{\nolinkurl{http://ascl.net/#1}}}
\providecommand{\doarXiv}[1]{\href{https://arxiv.org/abs/#1}{\nolinkurl{https://arxiv.org/abs/#1}}}

\bibitem[{Abadi {et~al.}(2015)Abadi, Agarwal, Barham, Brevdo, Chen, Citro,
  Corrado, Davis, Dean, Devin, Ghemawat, Goodfellow, Harp, Irving, Isard, Jia,
  Jozefowicz, Kaiser, Kudlur, Levenberg, Man\'{e}, Monga, Moore, Murray, Olah,
  Schuster, Shlens, Steiner, Sutskever, Talwar, Tucker, Vanhoucke, Vasudevan,
  Vi\'{e}gas, Vinyals, Warden, Wattenberg, Wicke, Yu, \&
  Zheng}]{tensorflow2015}
Abadi, M., Agarwal, A., Barham, P., {et~al.} 2015, {TensorFlow}: Large-Scale
  Machine Learning on Heterogeneous Systems.
\newblock \url{http://tensorflow.org/}

\bibitem[{{Astropy Collaboration} {et~al.}(2013){Astropy Collaboration},
  {Robitaille}, {Tollerud}, {Greenfield}, {Droettboom}, {Bray}, {Aldcroft},
  {Davis}, {Ginsburg}, {Price-Whelan}, {Kerzendorf}, {Conley}, {Crighton},
  {Barbary}, {Muna}, {Ferguson}, {Grollier}, {Parikh}, {Nair}, {Unther},
  {Deil}, {Woillez}, {Conseil}, {Kramer}, {Turner}, {Singer}, {Fox}, {Weaver},
  {Zabalza}, {Edwards}, {Azalee Bostroem}, {Burke}, {Casey}, {Crawford},
  {Dencheva}, {Ely}, {Jenness}, {Labrie}, {Lim}, {Pierfederici}, {Pontzen},
  {Ptak}, {Refsdal}, {Servillat}, \& {Streicher}}]{astropy}
{Astropy Collaboration}, {Robitaille}, T.~P., {Tollerud}, E.~J., {et~al.} 2013,
  \aap, 558, A33, \dodoi{10.1051/0004-6361/201322068}

\bibitem[{{Bandeira} {et~al.}(2017){Bandeira}, {Blum-Smith}, {Kileel}, {Perry},
  {Weed}, \& {Wein}}]{bandeira17}
{Bandeira}, A.~S., {Blum-Smith}, B., {Kileel}, J., {et~al.} 2017, arXiv
  e-prints, arXiv:1712.10163.
\newblock \doarXiv{1712.10163}

\bibitem[{{Bendek} {et~al.}(2018){Bendek}, {Tuthill}, {Guyon}, {Vasisht},
  {Belikov}, {Larkin}, {Beichman}, {Shao}, {Mamajek}, {Sirbu}, \&
  {Coyle}}]{bendek18}
{Bendek}, E., {Tuthill}, P., {Guyon}, O., {et~al.} 2018, in Society of
  Photo-Optical Instrumentation Engineers (SPIE) Conference Series, Vol. 10698,
  \procspie, 106980G, \dodoi{10.1117/12.2313919}

\bibitem[{{Bezanson} {et~al.}(2012){Bezanson}, {Karpinski}, {Shah}, \&
  {Edelman}}]{julia}
{Bezanson}, J., {Karpinski}, S., {Shah}, V.~B., \& {Edelman}, A. 2012, arXiv
  e-prints, arXiv:1209.5145.
\newblock \doarXiv{1209.5145}

\bibitem[{{Blackburn} {et~al.}(2020){Blackburn}, {Pesce}, {Johnson}, {Wielgus},
  {Chael}, {Christian}, \& {Doeleman}}]{blackburn20}
{Blackburn}, L., {Pesce}, D.~W., {Johnson}, M.~D., {et~al.} 2020, \apj, 894,
  31, \dodoi{10.3847/1538-4357/ab8469}

\bibitem[{{Bland-Hawthorn} \& {Leon-Saval}(2017)}]{jbh17}
{Bland-Hawthorn}, J., \& {Leon-Saval}, S.~G. 2017, Optics Express, 25, 15549,
  \dodoi{10.1364/OE.25.015549}

\bibitem[{{Bloemhof} {et~al.}(2001){Bloemhof}, {Dekany}, {Troy}, \&
  {Oppenheimer}}]{bloemhof01}
{Bloemhof}, E.~E., {Dekany}, R.~G., {Troy}, M., \& {Oppenheimer}, B.~R. 2001,
  \apjl, 558, L71, \dodoi{10.1086/323494}

\bibitem[{{Bracewell}(1978)}]{bracewell78}
{Bracewell}, R.~N. 1978, \nat, 274, 780, \dodoi{10.1038/274780a0}

\bibitem[{Bradbury {et~al.}(2018)Bradbury, Frostig, Hawkins, Johnson, Leary,
  Maclaurin, \& Wanderman-Milne}]{jax}
Bradbury, J., Frostig, R., Hawkins, P., {et~al.} 2018, {JAX}: composable
  transformations of {P}ython+{N}um{P}y programs, 0.1.55.
\newblock \url{http://github.com/google/jax}

\bibitem[{{Carlotti} {et~al.}(2011){Carlotti}, {Vanderbei}, \&
  {Kasdin}}]{carlotti11}
{Carlotti}, A., {Vanderbei}, R., \& {Kasdin}, N.~J. 2011, Optics Express, 19,
  26796, \dodoi{10.1364/OE.19.026796}

\bibitem[{{Ceau} {et~al.}(2019){Ceau}, {Mary}, {Greenbaum}, {Martinache},
  {Sivaramakrishnan}, {Laugier}, \& {N'Diaye}}]{ceau19}
{Ceau}, A., {Mary}, D., {Greenbaum}, A., {et~al.} 2019, arXiv e-prints.
\newblock \doarXiv{1908.03130}

\bibitem[{{Chael} {et~al.}(2018){Chael}, {Johnson}, {Bouman}, {Blackburn},
  {Akiyama}, \& {Narayan}}]{chael18}
{Chael}, A.~A., {Johnson}, M.~D., {Bouman}, K.~L., {et~al.} 2018, \apj, 857,
  23, \dodoi{10.3847/1538-4357/aab6a8}

\bibitem[{{Chaware} {et~al.}(2019){Chaware}, {Cooke}, {Kim}, \&
  {Horstmeyer}}]{chaware19}
{Chaware}, A., {Cooke}, C.~L., {Kim}, K., \& {Horstmeyer}, R. 2019, arXiv
  e-prints, arXiv:1910.10209.
\newblock \doarXiv{1910.10209}

\bibitem[{{Chianese} {et~al.}(2019){Chianese}, {Coogan}, {Hofma}, {Otten}, \&
  {Weniger}}]{chianese19}
{Chianese}, M., {Coogan}, A., {Hofma}, P., {Otten}, S., \& {Weniger}, C. 2019,
  arXiv e-prints, arXiv:1910.06157.
\newblock \doarXiv{1910.06157}

\bibitem[{{Currie} {et~al.}(2020){Currie}, {Pluzhnik}, {Guyon}, {Belikov},
  {Miller}, {Bos}, {Males}, {Sirbu}, {Bond}, {Frazin}, {Groff}, {Kern}, {Lozi},
  {Mazin}, {Nemati}, {Norris}, {Subedi}, \& {Will}}]{currie20}
{Currie}, T., {Pluzhnik}, E., {Guyon}, O., {et~al.} 2020, arXiv e-prints,
  arXiv:2007.14413.
\newblock \doarXiv{2007.14413}

\bibitem[{Czekala(2019)}]{czekala19}
Czekala, I. 2019, iancze/MPoL: Base version, v0.0.1,  Zenodo,
  \dodoi{10.5281/zenodo.3594082}

\bibitem[{{Doelman} {et~al.}(2018){Doelman}, {Tuthill}, {Norris}, {Wilby},
  {Por}, {Keller}, {Escuti}, \& {Snik}}]{doelman18}
{Doelman}, D.~S., {Tuthill}, P., {Norris}, B., {et~al.} 2018, in Society of
  Photo-Optical Instrumentation Engineers (SPIE) Conference Series, Vol. 10701,
  \procspie, 107010T, \dodoi{10.1117/12.2313547}

\bibitem[{Du {et~al.}(2020)Du, Nashed, Kandel, G{\"u}rsoy, \&
  Jacobsen}]{Dueaay3700}
Du, M., Nashed, Y. S.~G., Kandel, S., G{\"u}rsoy, D., \& Jacobsen, C. 2020,
  Science Advances, 6, \dodoi{10.1126/sciadv.aay3700}

\bibitem[{{Feinstein} {et~al.}(2019){Feinstein}, {Montet}, {Foreman-Mackey},
  {Bedell}, {Saunders}, {Bean}, {Christiansen}, {Hedges}, {Luger}, {Scolnic},
  \& {Cardoso}}]{eleanor}
{Feinstein}, A.~D., {Montet}, B.~T., {Foreman-Mackey}, D., {et~al.} 2019,
  \pasp, 131, 094502, \dodoi{10.1088/1538-3873/ab291c}

\bibitem[{{Fizeau}(1868)}]{fizeau1868}
{Fizeau}, H. 1868, Comptes Rendus de l'Acad\'{e}mie des Sciences, 66, 932

\bibitem[{{Foo} {et~al.}(2005){Foo}, {Palacios}, \& {Swartzland er}}]{foo05}
{Foo}, G., {Palacios}, D.~M., \& {Swartzland er}, Grover~A., J. 2005, Optics
  Letters, 30, 3308, \dodoi{10.1364/OL.30.003308}

\bibitem[{Foreman-Mackey(2016)}]{corner}
Foreman-Mackey, D. 2016, The Journal of Open Source Software, 24,
  \dodoi{10.21105/joss.00024}

\bibitem[{{Foreman-Mackey} {et~al.}(2013){Foreman-Mackey}, {Hogg}, {Lang}, \&
  {Goodman}}]{emcee}
{Foreman-Mackey}, D., {Hogg}, D.~W., {Lang}, D., \& {Goodman}, J. 2013, PASP,
  125, 306, \dodoi{10.1086/670067}

\bibitem[{{Guo} {et~al.}(2019){Guo}, {Barrett}, {Wang}, \& {Lvovsky}}]{guo19}
{Guo}, X., {Barrett}, T.~D., {Wang}, Z.~M., \& {Lvovsky}, A.~I. 2019, arXiv
  e-prints, arXiv:1912.12256.
\newblock \doarXiv{1912.12256}

\bibitem[{{Guyon}(2003)}]{guyon03}
{Guyon}, O. 2003, \aap, 404, 379, \dodoi{10.1051/0004-6361:20030457}

\bibitem[{{Guyon} {et~al.}(2012){Guyon}, {Bendek}, {Eisner}, {Angel}, {Woolf},
  {Milster}, {Ammons}, {Shao}, {Shaklan}, {Levine}, {Nemati}, {Pitman},
  {Woodruff}, \& {Belikov}}]{guyon12}
{Guyon}, O., {Bendek}, E.~A., {Eisner}, J.~A., {et~al.} 2012, \apjs, 200, 11,
  \dodoi{10.1088/0067-0049/200/2/11}

\bibitem[{Harris {et~al.}(2020)Harris, Millman, van~der Walt, Gommers,
  Virtanen, Cournapeau, Wieser, Taylor, Berg, Smith, Kern, Picus, Hoyer, van
  Kerkwijk, Brett, Haldane, del R{\'\i}o, Wiebe, Peterson, G{\'e}rard-Marchant,
  Sheppard, Reddy, Weckesser, Abbasi, Gohlke, \& Oliphant}]{numpy}
Harris, C.~R., Millman, K.~J., van~der Walt, S.~J., {et~al.} 2020, Nature, 585,
  357

\bibitem[{Hughes {et~al.}(2018)Hughes, Minkov, Williamson, \& Fan}]{hughes18}
Hughes, T.~W., Minkov, M., Williamson, I. A.~D., \& Fan, S. 2018, ACS
  Photonics, 5, 4781, \dodoi{10.1021/acsphotonics.8b01522}

\bibitem[{Hughes {et~al.}(2019)Hughes, Williamson, Minkov, \& Fan}]{hughes19}
Hughes, T.~W., Williamson, I. A.~D., Minkov, M., \& Fan, S. 2019, Science
  Advances, 5, \dodoi{10.1126/sciadv.aay6946}

\bibitem[{Hunter(2007)}]{matplotlib}
Hunter, J.~D. 2007, Computing In Science \& Engineering, 9, 90

\bibitem[{{Ireland}(2013)}]{ireland13}
{Ireland}, M.~J. 2013, \mnras, 433, 1718, \dodoi{10.1093/mnras/stt859}

\bibitem[{{Jennison}(1958)}]{jennison58}
{Jennison}, R.~C. 1958, \mnras, 118, 276, \dodoi{10.1093/mnras/118.3.276}

\bibitem[{Jones {et~al.}(2001)Jones, Oliphant, Peterson, \& Others}]{scipy}
Jones, E., Oliphant, T., Peterson, P., \& Others. 2001, {SciPy}: Open source
  scientific tools for Python.
\newblock \url{http://www.scipy.org/}

\bibitem[{{Jurling} \& {Fienup}(2014)}]{jurling14}
{Jurling}, A.~S., \& {Fienup}, J.~R. 2014, Journal of the Optical Society of
  America A, 31, 1348, \dodoi{10.1364/JOSAA.31.001348}

\bibitem[{{Kammerer} {et~al.}(2019){Kammerer}, {Ireland}, {Martinache}, \&
  {Girard}}]{kammerer19}
{Kammerer}, J., {Ireland}, M.~J., {Martinache}, F., \& {Girard}, J.~H. 2019,
  \mnras, 486, 639, \dodoi{10.1093/mnras/stz882}

\bibitem[{{Kandel} {et~al.}(2019){Kandel}, {Maddali}, {Allain}, {Hruszkewycz},
  {Jacobsen}, \& {Nashed}}]{kandel19}
{Kandel}, S., {Maddali}, S., {Allain}, M., {et~al.} 2019, Optics Express, 27,
  18653, \dodoi{10.1364/OE.27.018653}

\bibitem[{{Lafreni{\`e}re} {et~al.}(2007){Lafreni{\`e}re}, {Marois}, {Doyon},
  {Nadeau}, \& {Artigau}}]{lafreniere07}
{Lafreni{\`e}re}, D., {Marois}, C., {Doyon}, R., {Nadeau}, D., \& {Artigau},
  {\'E}. 2007, \apj, 660, 770, \dodoi{10.1086/513180}

\bibitem[{{Lallo} {et~al.}(2006){Lallo}, {Makidon}, {Casertano}, \&
  {Krist}}]{lallo06}
{Lallo}, M.~D., {Makidon}, R.~B., {Casertano}, S., \& {Krist}, J.~E. 2006, in
  Society of Photo-Optical Instrumentation Engineers (SPIE) Conference Series,
  Vol. 6270, Society of Photo-Optical Instrumentation Engineers (SPIE)
  Conference Series, 62701N, \dodoi{10.1117/12.672040}

\bibitem[{{Lannes}(1991)}]{lannes1991}
{Lannes}, A. 1991, Inverse Problems, 7, 261, \dodoi{10.1088/0266-5611/7/2/009}

\bibitem[{{Laugier} \& {Martinache}(2019)}]{laugier19b}
{Laugier}, R., \& {Martinache}, F. 2019, in SF2A-2019: Proceedings of the
  Annual meeting of the French Society of Astronomy and Astrophysics, Di

\bibitem[{{Laugier} {et~al.}(2019){Laugier}, {Martinache}, {Ceau}, {Mary},
  {N'Diaye}, \& {Beuzit}}]{laugier19}
{Laugier}, R., {Martinache}, F., {Ceau}, A., {et~al.} 2019, \aap, 623, A164,
  \dodoi{10.1051/0004-6361/201834387}

\bibitem[{{Laugier} {et~al.}(2020){Laugier}, {Martinache}, {Cvetojevic},
  {Mary}, {Ceau}, {N'Diaye}, {Kammerer}, {Lozi}, {Guyon}, \&
  {Lopez}}]{laugier20}
{Laugier}, R., {Martinache}, F., {Cvetojevic}, N., {et~al.} 2020, \aap, 636,
  A21, \dodoi{10.1051/0004-6361/201937121}

\bibitem[{{LeCun} {et~al.}(2015){LeCun}, {Bengio}, \& {Hinton}}]{lecun15}
{LeCun}, Y., {Bengio}, Y., \& {Hinton}, G. 2015, \nat, 521, 436,
  \dodoi{10.1038/nature14539}

\bibitem[{LeCun {et~al.}(1988)LeCun, Touresky, Hinton, \&
  Sejnowski}]{lecun1988theoretical}
LeCun, Y., Touresky, D., Hinton, G., \& Sejnowski, T. 1988, in Proceedings of
  the 1988 connectionist models summer school, Vol.~1, CMU, Pittsburgh, Pa:
  Morgan Kaufmann, 21--28

\bibitem[{Linnainmaa(1970)}]{linnainmaa1970}
Linnainmaa, S. 1970, Master's Thesis (in Finnish), Univ. Helsinki, 6

\bibitem[{{Lyot}(1930)}]{lyot30}
{Lyot}, B. 1930, Bulletin Astronomique, 6, 305

\bibitem[{Maclaurin {et~al.}(2015)Maclaurin, Duvenaud, \& Adams}]{autograd}
Maclaurin, D., Duvenaud, D., \& Adams, R.~P. 2015, in ICML 2015 AutoML
  Workshop, Vol. 238

\bibitem[{{Malbet} {et~al.}(1995){Malbet}, {Yu}, \& {Shao}}]{malbet95}
{Malbet}, F., {Yu}, J.~W., \& {Shao}, M. 1995, \pasp, 107, 386,
  \dodoi{10.1086/133563}

\bibitem[{{Marois} {et~al.}(2006){Marois}, {Lafreni{\`e}re}, {Doyon},
  {Macintosh}, \& {Nadeau}}]{marois06}
{Marois}, C., {Lafreni{\`e}re}, D., {Doyon}, R., {Macintosh}, B., \& {Nadeau},
  D. 2006, \apj, 641, 556, \dodoi{10.1086/500401}

\bibitem[{{Martinache}(2010)}]{martinache10}
{Martinache}, F. 2010, \apj, 724, 464, \dodoi{10.1088/0004-637X/724/1/464}

\bibitem[{{Martinache}(2011)}]{martinache11}
{Martinache}, F. 2011, in Society of Photo-Optical Instrumentation Engineers
  (SPIE) Conference Series, Vol. 8151, Society of Photo-Optical Instrumentation
  Engineers (SPIE) Conference Series, \dodoi{10.1117/12.894319}

\bibitem[{{Martinache}(2013)}]{martinache13}
---. 2013, \pasp, 125, 422, \dodoi{10.1086/670670}

\bibitem[{{Martinache}(2016)}]{martinache16}
---. 2016, Society of Photo-Optical Instrumentation Engineers (SPIE) Conference
  Series, Vol. 9907, {Spectrally dispersed Fourier-phase analysis for redundant
  apertures}, 990712, \dodoi{10.1117/12.2233395}

\bibitem[{Martinache(2018)}]{martinache_habilitation}
Martinache, F. 2018, `Repousser les limites de la diffraction pour l'astronomie
  \`a haute r\'esolution angulaire', Habilitation, Universit\'e de Nice Sophia
  Antipolis.
\newblock \url{http://frantzmartinache.eu/static/share/hdr_no_papers.pdf}

\bibitem[{{Martinache} {et~al.}(2020){Martinache}, {Ceau}, {Laugier},
  {Kammerer}, {N'Diaye}, {Mary}, {Cvetojevic}, \& {Lopez}}]{martinache20}
{Martinache}, F., {Ceau}, A., {Laugier}, R., {et~al.} 2020, arXiv e-prints,
  arXiv:2003.02032.
\newblock \doarXiv{2003.02032}

\bibitem[{{Martinache} \& {Ireland}(2018)}]{martinache18}
{Martinache}, F., \& {Ireland}, M.~J. 2018, \aap, 619, A87,
  \dodoi{10.1051/0004-6361/201832847}

\bibitem[{{Martinache} {et~al.}(2016){Martinache}, {Jovanovic}, \&
  {Guyon}}]{martinache16b}
{Martinache}, F., {Jovanovic}, N., \& {Guyon}, O. 2016, \aap, 593, A33,
  \dodoi{10.1051/0004-6361/201628496}

\bibitem[{Metropolis {et~al.}(1953)Metropolis, Rosenbluth, Rosenbluth, Teller,
  \& Teller}]{metropolis53}
Metropolis, N., Rosenbluth, A.~W., Rosenbluth, M.~N., Teller, A.~H., \& Teller,
  E. 1953, The Journal of Chemical Physics, 21, 1087, \dodoi{10.1063/1.1699114}

\bibitem[{{Minardi} {et~al.}(2020){Minardi}, {Harris}, \&
  {Labadie}}]{minardi20}
{Minardi}, S., {Harris}, R., \& {Labadie}, L. 2020, arXiv e-prints,
  arXiv:2003.12485.
\newblock \doarXiv{2003.12485}

\bibitem[{{Morningstar} {et~al.}(2018){Morningstar}, {Hezaveh}, {Perreault
  Levasseur}, {Blandford}, {Marshall}, {Putzky}, \& {Wechsler}}]{morningstar18}
{Morningstar}, W.~R., {Hezaveh}, Y.~D., {Perreault Levasseur}, L., {et~al.}
  2018, arXiv e-prints, arXiv:1808.00011.
\newblock \doarXiv{1808.00011}

\bibitem[{{Morningstar} {et~al.}(2019){Morningstar}, {Perreault Levasseur},
  {Hezaveh}, {Blandford}, {Marshall}, {Putzky}, {Rueter}, {Wechsler}, \&
  {Welling}}]{morningstar19}
{Morningstar}, W.~R., {Perreault Levasseur}, L., {Hezaveh}, Y.~D., {et~al.}
  2019, \apj, 883, 14, \dodoi{10.3847/1538-4357/ab35d7}

\bibitem[{Muthumbi {et~al.}(2019)Muthumbi, Chaware, Kim, Zhou, Konda, Chen,
  Judkewitz, Erdmann, Kappes, \& Horstmeyer}]{muthumbi19}
Muthumbi, A., Chaware, A., Kim, K., {et~al.} 2019, Biomed. Opt. Express, 10,
  6351, \dodoi{10.1364/BOE.10.006351}

\bibitem[{{Nardiello} {et~al.}(2019){Nardiello}, {Borsato}, {Piotto},
  {Colombo}, {Manthopoulou}, {Bedin}, {Granata}, {Lacedelli}, {Libralato},
  {Malavolta}, {Montalto}, \& {Nascimbeni}}]{nardiello19}
{Nardiello}, D., {Borsato}, L., {Piotto}, G., {et~al.} 2019, \mnras, 490, 3806,
  \dodoi{10.1093/mnras/stz2878}

\bibitem[{{Nashed} {et~al.}(2019){Nashed}, {Kandel}, {Du}, \&
  {Jacobsen}}]{nashed19}
{Nashed}, Y. S.~G., {Kandel}, S., {Du}, M., \& {Jacobsen}, C. 2019, Microscopy
  and Microanalysis, 25, 62, \dodoi{10.1017/S1431927619001041}

\bibitem[{{Norris} {et~al.}(2014){Norris}, {Tuthill}, {Jovanovic}, {Schworer},
  {Guyon}, {Stewart}, \& {Martinache}}]{vampires}
{Norris}, B., {Tuthill}, P., {Jovanovic}, N., {et~al.} 2014, arXiv e-prints,
  arXiv:1405.7426.
\newblock \doarXiv{1405.7426}

\bibitem[{{Paine} \& {Fienup}(2019)}]{paine19}
{Paine}, S.~W., \& {Fienup}, J.~R. 2019, in Society of Photo-Optical
  Instrumentation Engineers (SPIE) Conference Series, Vol. 10980, \procspie,
  109800T, \dodoi{10.1117/12.2519884}

\bibitem[{{Paszke} {et~al.}(2019){Paszke}, {Gross}, {Massa}, {Lerer},
  {Bradbury}, {Chanan}, {Killeen}, {Lin}, {Gimelshein}, {Antiga}, {Desmaison},
  {K{\"o}pf}, {Yang}, {DeVito}, {Raison}, {Tejani}, {Chilamkurthy}, {Steiner},
  {Fang}, {Bai}, \& {Chintala}}]{pytorch}
{Paszke}, A., {Gross}, S., {Massa}, F., {et~al.} 2019, arXiv e-prints,
  arXiv:1912.01703.
\newblock \doarXiv{1912.01703}

\bibitem[{P\'erez \& Granger(2007)}]{ipython}
P\'erez, F., \& Granger, B.~E. 2007, Computing in Science and Engineering, 9,
  21, \dodoi{10.1109/MCSE.2007.53}

\bibitem[{{Perrin} {et~al.}(2003){Perrin}, {Sivaramakrishnan}, {Makidon},
  {Oppenheimer}, \& {Graham}}]{perrin03}
{Perrin}, M.~D., {Sivaramakrishnan}, A., {Makidon}, R.~B., {Oppenheimer},
  B.~R., \& {Graham}, J.~R. 2003, \apj, 596, 702, \dodoi{10.1086/377689}

\bibitem[{{Perrin} {et~al.}(2012{\natexlab{a}}){Perrin}, {Soummer}, {Elliott},
  {Lallo}, \& {Sivaramakrishnan}}]{poppy}
{Perrin}, M.~D., {Soummer}, R., {Elliott}, E.~M., {Lallo}, M.~D., \&
  {Sivaramakrishnan}, A. 2012{\natexlab{a}}, Society of Photo-Optical
  Instrumentation Engineers (SPIE) Conference Series, Vol. 8442, {Simulating
  point spread functions for the James Webb Space Telescope with WebbPSF},
  84423D, \dodoi{10.1117/12.925230}

\bibitem[{{Perrin} {et~al.}(2012{\natexlab{b}}){Perrin}, {Soummer}, {Elliott},
  {Lallo}, \& {Sivaramakrishnan}}]{webbpsf}
---. 2012{\natexlab{b}}, Society of Photo-Optical Instrumentation Engineers
  (SPIE) Conference Series, Vol. 8442, {Simulating point spread functions for
  the James Webb Space Telescope with WebbPSF}, 84423D,
  \dodoi{10.1117/12.925230}

\bibitem[{{Perrin} {et~al.}(2015){Perrin}, {Duchene}, {Millar-Blanchaer},
  {Fitzgerald}, {Graham}, {Wiktorowicz}, {Kalas}, {Macintosh}, {Bauman},
  {Cardwell}, {Chilcote}, {De Rosa}, {Dillon}, {Doyon}, {Dunn}, {Erikson},
  {Gavel}, {Goodsell}, {Hartung}, {Hibon}, {Ingraham}, {Kerley}, {Konapacky},
  {Larkin}, {Maire}, {Marchis}, {Marois}, {Mittal}, {Morzinski}, {Oppenheimer},
  {Palmer}, {Patience}, {Poyneer}, {Pueyo}, {Rantakyr{\"o}}, {Sadakuni},
  {Saddlemyer}, {Savransky}, {Soummer}, {Sivaramakrishnan}, {Song}, {Thomas},
  {Wallace}, {Wang}, \& {Wolff}}]{gpipol}
{Perrin}, M.~D., {Duchene}, G., {Millar-Blanchaer}, M., {et~al.} 2015, \apj,
  799, 182, \dodoi{10.1088/0004-637X/799/2/182}

\bibitem[{{Pope} {et~al.}(2014{\natexlab{a}}){Pope}, {Cvetojevic}, {Cheetham},
  {Martinache}, {Norris}, \& {Tuthill}}]{pope14}
{Pope}, B., {Cvetojevic}, N., {Cheetham}, A., {et~al.} 2014{\natexlab{a}},
  \mnras, 440, 125, \dodoi{10.1093/mnras/stu218}

\bibitem[{{Pope} {et~al.}(2013){Pope}, {Martinache}, \& {Tuthill}}]{pope13}
{Pope}, B., {Martinache}, F., \& {Tuthill}, P. 2013, \apj, 767, 110,
  \dodoi{10.1088/0004-637X/767/2/110}

\bibitem[{{Pope} {et~al.}(2014{\natexlab{b}}){Pope}, {Thatte}, {Burruss},
  {Tecza}, {Clarke}, \& {Cotter}}]{swift}
{Pope}, B., {Thatte}, N., {Burruss}, R., {et~al.} 2014{\natexlab{b}}, Society
  of Photo-Optical Instrumentation Engineers (SPIE) Conference Series, Vol.
  9148, {Wavefront sensing from the image domain with the Oxford-SWIFT integral
  field spectrograph}, 914859, \dodoi{10.1117/12.2055334}

\bibitem[{{Pope} {et~al.}(2016){Pope}, {Tuthill}, {Hinkley}, {Ireland },
  {Greenbaum}, {Latyshev}, {Monnier}, \& {Martinache}}]{palomar}
{Pope}, B., {Tuthill}, P., {Hinkley}, S., {et~al.} 2016, \mnras, 455, 1647,
  \dodoi{10.1093/mnras/stv2442}

\bibitem[{{Pope}(2016)}]{pope16}
{Pope}, B. J.~S. 2016, \mnras, 463, 3573, \dodoi{10.1093/mnras/stw2215}

\bibitem[{{Pope} {et~al.}(2019){Pope}, {White}, {Farr}, {Yu}, {Greklek-McKeon},
  {Huber}, {Aerts}, {Aigrain}, {Bedding}, {Boyajian}, {Creevey}, \&
  {Hogg}}]{pope19}
{Pope}, B. J.~S., {White}, T.~R., {Farr}, W.~M., {et~al.} 2019, \apjs, 245, 8,
  \dodoi{10.3847/1538-4365/ab3d29}

\bibitem[{{Pueyo}(2016)}]{pueyo16}
{Pueyo}, L. 2016, \apj, 824, 117, \dodoi{10.3847/0004-637X/824/2/117}

\bibitem[{{Rieke} {et~al.}(2015){Rieke}, {Ressler}, {Morrison}, {Bergeron},
  {Bouchet}, {Garc{\'\i}a-Mar{\'\i}n}, {Greene}, {Regan}, {Sukhatme}, \&
  {Walker}}]{rieke15}
{Rieke}, G.~H., {Ressler}, M.~E., {Morrison}, J.~E., {et~al.} 2015, \pasp, 127,
  665, \dodoi{10.1086/682257}

\bibitem[{{Riggs} {et~al.}(2018){Riggs}, {Ruane}, {Sidick}, {Coker}, {Kern}, \&
  {Shaklan}}]{falco}
{Riggs}, A.~J.~E., {Ruane}, G., {Sidick}, E., {et~al.} 2018, in Society of
  Photo-Optical Instrumentation Engineers (SPIE) Conference Series, Vol. 10698,
  \procspie, 106982V, \dodoi{10.1117/12.2313812}

\bibitem[{{Sallum} \& {Skemer}(2019)}]{sallum19a}
{Sallum}, S., \& {Skemer}, A. 2019, Journal of Astronomical Telescopes,
  Instruments, and Systems, 5, 018001, \dodoi{10.1117/1.JATIS.5.1.018001}

\bibitem[{{Sallum} {et~al.}(2015){Sallum}, {Follette}, {Eisner}, {Close},
  {Hinz}, {Kratter}, {Males}, {Skemer}, {Macintosh}, {Tuthill}, {Bailey},
  {Defr{\`e}re}, {Morzinski}, {Rodigas}, {Spalding}, {Vaz}, \&
  {Weinberger}}]{sallum15}
{Sallum}, S., {Follette}, K.~B., {Eisner}, J.~A., {et~al.} 2015, \nat, 527,
  342, \dodoi{10.1038/nature15761}

\bibitem[{{Sallum} {et~al.}(2019){Sallum}, {Skemer}, {Eisner}, {van der Marel},
  {Sheehan}, {Close}, {Ireland}, {Males}, {Morzinski}, {Bailey}, {Briguglio},
  \& {Puglisi}}]{sallum19b}
{Sallum}, S., {Skemer}, A., {Eisner}, J., {et~al.} 2019, arXiv e-prints.
\newblock \doarXiv{1908.07427}

\bibitem[{{Schmid} {et~al.}(2018){Schmid}, {Bazzon}, {Roelfsema}, {Mouillet},
  {Milli}, {Menard}, {Gisler}, {Hunziker}, {Pragt}, {Dominik}, {Boccaletti},
  {Ginski}, {Abe}, {Antoniucci}, {Avenhaus}, {Baruffolo}, {Baudoz}, {Beuzit},
  {Carbillet}, {Chauvin}, {Claudi}, {Costille}, {Daban}, {de Haan}, {Desidera},
  {Dohlen}, {Downing}, {Elswijk}, {Engler}, {Feldt}, {Fusco}, {Girard},
  {Gratton}, {Hanenburg}, {Henning}, {Hubin}, {Joos}, {Kasper}, {Keller},
  {Langlois}, {Lagadec}, {Martinez}, {Mulder}, {Pavlov}, {Podio}, {Puget},
  {Quanz}, {Rigal}, {Salasnich}, {Sauvage}, {Schuil}, {Siebenmorgen}, {Sissa},
  {Snik}, {Suarez}, {Thalmann}, {Turatto}, {Udry}, {van Duin}, {van Holstein},
  {Vigan}, \& {Wildi}}]{zimpol}
{Schmid}, H.~M., {Bazzon}, A., {Roelfsema}, R., {et~al.} 2018, \aap, 619, A9,
  \dodoi{10.1051/0004-6361/201833620}

\bibitem[{Sitzmann {et~al.}(2018)Sitzmann, Diamond, Peng, Dun, Boyd, Heidrich,
  Heide, \& Wetzstein}]{sitzmann2018}
Sitzmann, V., Diamond, S., Peng, Y., {et~al.} 2018, ACM Transactions on
  Graphics (TOG), 37, 114

\bibitem[{{Sivaramakrishnan} {et~al.}(2002){Sivaramakrishnan}, {Lloyd},
  {Hodge}, \& {Macintosh}}]{anand02}
{Sivaramakrishnan}, A., {Lloyd}, J.~P., {Hodge}, P.~E., \& {Macintosh}, B.~A.
  2002, \apjl, 581, L59, \dodoi{10.1086/345826}

\bibitem[{{Soummer} {et~al.}(2012){Soummer}, {Pueyo}, \& {Larkin}}]{soummer12}
{Soummer}, R., {Pueyo}, L., \& {Larkin}, J. 2012, \apjl, 755, L28,
  \dodoi{10.1088/2041-8205/755/2/L28}

\bibitem[{{Soummer} {et~al.}(2007){Soummer}, {Pueyo}, {Sivaramakrishnan}, \&
  {Vand erbei}}]{soummer07}
{Soummer}, R., {Pueyo}, L., {Sivaramakrishnan}, A., \& {Vand erbei}, R.~J.
  2007, Optics Express, 15, 15935, \dodoi{10.1364/OE.15.015935}

\bibitem[{{Sparks} \& {Ford}(2002)}]{sparks02}
{Sparks}, W.~B., \& {Ford}, H.~C. 2002, \apj, 578, 543, \dodoi{10.1086/342401}

\bibitem[{{Sutin}(2016)}]{sutin16}
{Sutin}, B.~M. 2016, Society of Photo-Optical Instrumentation Engineers (SPIE)
  Conference Series, Vol. 9911, {An optical toolbox for astronomical
  instrumentation}, 99112J, \dodoi{10.1117/12.2233677}

\bibitem[{{Theano Development Team}(2016)}]{theano}
{Theano Development Team}. 2016, arXiv e-prints, abs/1605.02688.
\newblock \url{http://arxiv.org/abs/1605.02688}

\bibitem[{{Tuthill} {et~al.}(2018){Tuthill}, {Bendek}, {Guyon}, {Horton},
  {Jeffries}, {Jovanovic}, {Klupar}, {Larkin}, {Norris}, {Pope}, \&
  {Shao}}]{tuthill18}
{Tuthill}, P., {Bendek}, E., {Guyon}, O., {et~al.} 2018, in Society of
  Photo-Optical Instrumentation Engineers (SPIE) Conference Series, Vol. 10701,
  \procspie, 107011J, \dodoi{10.1117/12.2313269}

\bibitem[{{Twiss} {et~al.}(1960){Twiss}, {Carter}, \& {Little}}]{twiss60}
{Twiss}, R.~Q., {Carter}, A.~W.~L., \& {Little}, A.~G. 1960, The Observatory,
  80, 153

\bibitem[{{van Cittert}(1934)}]{vc34}
{van Cittert}, P.~H. 1934, Physica, 1, 201,
  \dodoi{10.1016/S0031-8914(34)90026-4}

\bibitem[{Werner {et~al.}(2012)Werner, Hillenbrand, Hoffmann, Sinzinger,
  {et~al.}}]{werner2012}
Werner, J., Hillenbrand, M., Hoffmann, A., Sinzinger, S., {et~al.} 2012,
  Schedae Informaticae, 2012, 169

\bibitem[{{White} {et~al.}(2017){White}, {Pope}, {Antoci}, {P{\'a}pics},
  {Aerts}, {Gies}, {Gordon}, {Huber}, {Schaefer}, {Aigrain}, {Albrecht},
  {Barclay}, {Barentsen}, {Beck}, {Bedding}, {Fredslund Andersen}, {Grundahl},
  {Howell}, {Ireland}, {Murphy}, {Nielsen}, {Silva Aguirre}, \&
  {Tuthill}}]{white17}
{White}, T.~R., {Pope}, B.~J.~S., {Antoci}, V., {et~al.} 2017, \mnras, 471,
  2882, \dodoi{10.1093/mnras/stx1050}

\bibitem[{Williams {et~al.}(2006)Williams, Quiney, Dhal, Tran, Nugent, Peele,
  Paterson, \& de~Jonge}]{williams2006}
Williams, G.~J., Quiney, H.~M., Dhal, B.~B., {et~al.} 2006, Phys. Rev. Lett.,
  97, 025506, \dodoi{10.1103/PhysRevLett.97.025506}

\bibitem[{{Zernike}(1938)}]{zernike38}
{Zernike}, F. 1938, Physica, 5, 785, \dodoi{10.1016/S0031-8914(38)80203-2}

\bibitem[{{Zernike}(1934)}]{zernike34}
{Zernike}, v.~F. 1934, Physica, 1, 689, \dodoi{10.1016/S0031-8914(34)80259-5}

\end{thebibliography}



\end{document}